\documentclass{article}

\PassOptionsToPackage{round}{natbib}
\usepackage[preprint]{tackling_climate_workshop_style}

\usepackage[utf8]{inputenc} 
\usepackage[T1]{fontenc}    
\usepackage{hyperref}       
\usepackage{url}            
\usepackage{booktabs}       
\usepackage{amsfonts}       
\usepackage{nicefrac}       
\usepackage{microtype}      
\usepackage{xcolor}
\usepackage{graphicx}
\usepackage{xfrac}

\RequirePackage{doi}

\usepackage{color, colortbl}
\definecolor{darkblue}{rgb}{0.0,0.0,0.65}
\definecolor{darkred}{rgb}{0.68,0.05,0.0}
\definecolor{darkgreen}{rgb}{0.0,0.29,0.29}
\definecolor{darkpurple}{rgb}{0.47,0.09,0.29}
\definecolor{myblue}{rgb}{0.1380392156862745, 0.3027450980392157, 0.6274509803921569}
\definecolor{myred}{rgb}{0.6235294117647059, 0.13725490196078433, 0.0196078431372549} 
\hypersetup{
  colorlinks = true,
  citecolor  = myblue,
  linkcolor  = myred,
  filecolor  = darkblue,
  urlcolor   = darkblue,
 }

\usepackage{tikz, }
\usetikzlibrary{shapes,arrows,fit}

\title{ACE2: Accurately learning subseasonal to decadal atmospheric variability and forced responses}

\author{%
  Oliver Watt-Meyer\thanks{\texttt{oliverwm@allenai.org; $\,$\href{https://github.com/ai2cm/ace}{https://github.com/ai2cm/ace}}}, Brian Henn,  Jeremy McGibbon, Spencer K. Clark \\
  \textbf{Anna Kwa, W. Andre Perkins, Elynn Wu, Christopher S. Bretherton}
  \\
  Allen Institute for Artificial Intelligence (Ai2), Seattle, WA, USA
  \\
  \\
  \textbf{Lucas Harris}
  \\
  Geophysical Fluid Dynamics Laboratory, NOAA, Princeton, NJ, USA
}

\begin{document}

\maketitle

\begin{abstract}
  Existing machine learning models of weather variability are not formulated to enable assessment of their response to varying external boundary conditions such as sea surface temperature and greenhouse gases. Here we present ACE2 (Ai2 Climate Emulator version 2) and its application to reproducing atmospheric variability over the past 80 years on timescales from days to decades. ACE2 is a 450M-parameter autoregressive machine learning emulator, operating with 6-hour temporal resolution, 1° horizontal resolution and eight vertical layers. It exactly conserves global dry air mass and moisture and can be stepped forward stably for arbitrarily many steps with a throughput of about 1500 simulated years per wall clock day. ACE2 generates emergent phenomena such as tropical cyclones, the Madden Julian Oscillation, and sudden stratospheric warmings. Furthermore, it accurately reproduces the atmospheric response to El Ni\~no variability and global trends of temperature over the past 80 years. However, its sensitivities to separately changing sea surface temperature and carbon dioxide are not entirely realistic.
\end{abstract}

\section{Introduction}
Machine learning offers an avenue to accelerate existing climate models by orders of magnitude. This acceleration is achieved by running efficiently on GPU hardware and by taking relatively long time steps, enabled by the lack of stability constraints that accompany traditional numerical methods. This increased efficiency has the potential to dramatically accelerate research tasks requiring many years of simulation. For example, it would enable easier exploration of large ensembles and rare events \citep{Kay2015, Manesh2024} and allow accurate separation of forced response versus internal variability \citep{Milinski2020}. It would permit the lengthy simulations necessary for the study of paleoclimate with more realistic models than intermediate complexity models \citep{Claussen2002}. Finally, it would enable easy interpolation between wide range of climate change scenarios \citep{WatsonParris2022}. The cheap cost of inference and ability to run on consumer hardware opens the door of running climate models to a wider range of users. In addition to acceleration, a machine-learning based climate model emulator is differentiable, making it immediately useful for data assimilation applications \citep{Brajard2020,Hatfield2021,Perkins2021}.

The extent to which machine learning will lead to more accurate climate models remains to be seen. While machine learning has demonstrated an ability to improve weather prediction accuracy \citep[e.g.][]{PanguWeather,Lam2023,Price2023,Chen2024,Kochkov2024}, the typical goal of climate prediction is to forecast previously unseen conditions, for example the expected global warming from a doubling of CO$_2$ concentration. Out-of-sample generalization is a fundamental challenge for machine learning, potentially necessitating the use of physics-based priors \citep[e.g.][]{Kochov2021,Beucler2024} and the training of machine learning based climate emulators on output from physics-based numerical models \citep{Clark2022}. In this study we focus on emulating the climate of the historical period 1940--2020, including variability and trends. We demonstrate that our emulator can be skillfully trained on the ERA5 reanalysis \citep{Hersbach2020} or on an AMIP-style \citep{Eyring2016} historical simulation with GFDL's SHiELD model \citep{Harris2020}. SHiELD can also simulate perturbed climates, as would be needed to train an emulator that could be expected to simulate long-term climate change.

For this work, we use the Ai2 Climate Emulator version 2 (ACE2; see \href{https://github.com/ai2cm/ace}{https://github.com/ai2cm/ace}), a significant update to the ACE atmospheric model emulator described in \cite{WattMeyer2023} and \cite{Duncan2024}. Briefly, the emulator operates at 1° horizontal resolution with eight terrain-following vertical layers. It is initialized from a snapshot of atmospheric temperature, humidity and winds and can stably integrate forward an arbitrary number of 6-hour time steps with a user-specified sea surface temperature (SST) boundary condition. The main methodological advances of ACE2 over version 1 of ACE are: 1) addition of CO$_2$ as a forcing variable, 2) ability to emulate observed atmospheric trends of the preceding 80 years and 3) the exact conservation of dry air mass and atmospheric moisture in ACE2 simulations. In addition, ACE2 is trained on two datasets to demonstrate its general applicability: first on an AMIP-style \citep{Eyring2016} simulation with GFDL's SHiELD model \citep{Harris2020} and second on the ERA5 reanalysis \citep{Hersbach2020}.

This study provides a more multifaceted evaluation of ACE2 than in our prior work on ACE, which only used annually-repeating climatological SSTs. We show ACE2's accurate atmospheric response to El Ni\~no variability as well as the long-term trends and interannual variability of global mean temperature and total water path. The ERA5-trained model allows evaluation of weather forecast skill and of phenomena such as tropical cyclones and the Madden Julian Oscillation, which are less well represented in the relatively coarse atmospheric models previously used for training ACE.

Related work includes NeuralGCM \citep{Kochkov2024} which showed some 30-year simulations with reasonable trends and low climate biases. However about one third of NeuralGCM's  simulations went unstable before reaching 30 years, which limits its current applicability to climate prediction. Atmospheric emulators with long-term stability trained on ERA5 \citep[e.g.][]{Karlbauer2024,Guan2024,Cresswell2024} and atmospheric model output \citep{WattMeyer2023,Duncan2024,Salva2024} have been reported, but none to date demonstrate the ability to accurately respond to the changing external forcing of the atmosphere over the last 80 years.


\section{Results}
\label{sec:results}

\subsection{Training period evaluation}

We present ACE2 model evaluations initialized in January 1940 and run forward for 81 years through December 2020, spanning nearly the full period  of ERA5 and SHiELD data. Although this period overlaps with the training data, which covers 1940-1995 and 2011-2019 (see Methods), ACE2 is only trained to predict two 6-hourly time steps ahead, and so the long autoregressive rollouts shown here demonstrate ACE2's ability to run stably and respond to long-term forcing. We evaluate ACE2's inference performance on a held out 10-year test period in Section \ref{subsubsec:holdout_climate_skill}. 

\begin{figure}[t]
    \centering
    \includegraphics[width=\textwidth]{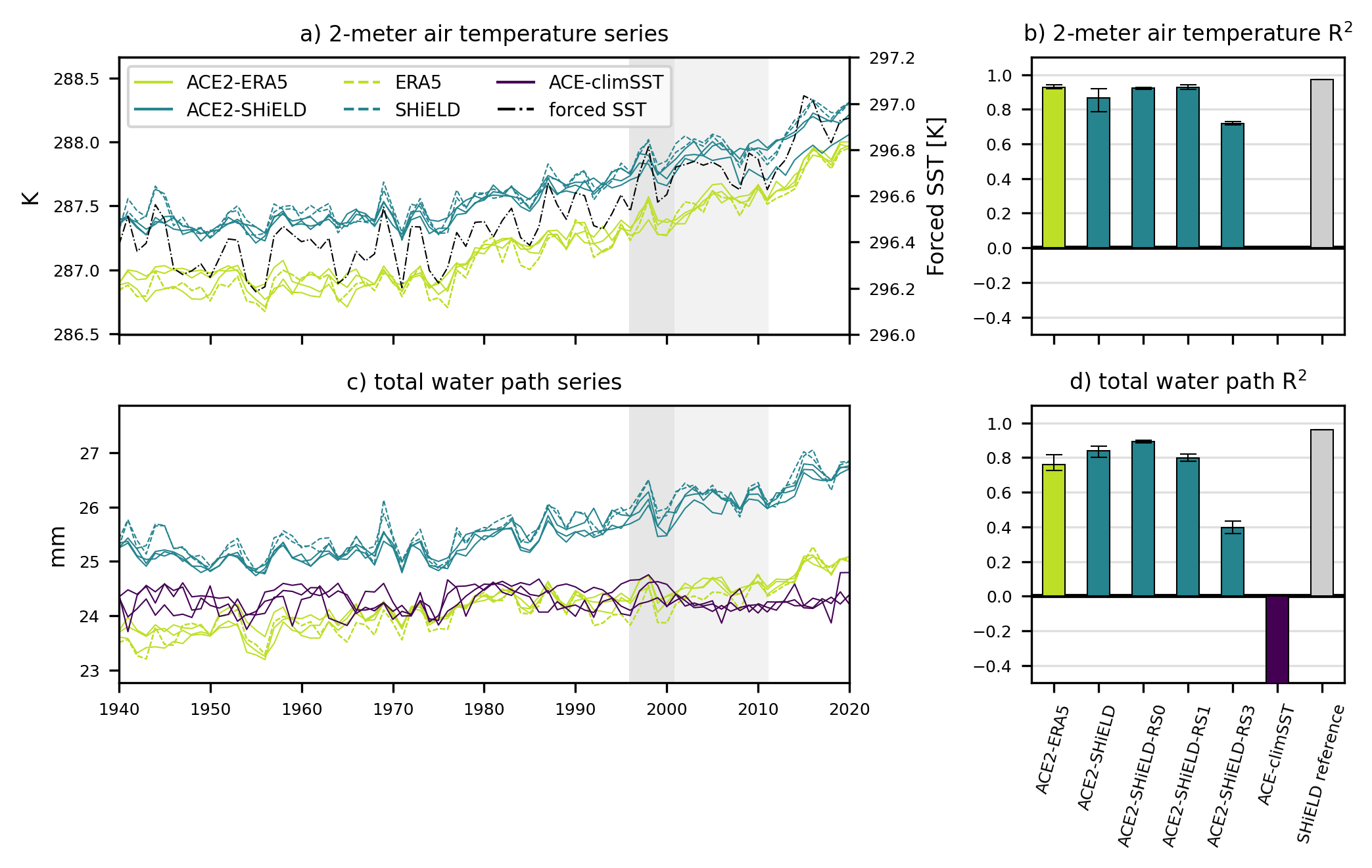}
    \caption{Global- and annual-mean series for a) 2-meter air temperature and c) total water path over 81-year evaluations of ACE2-ERA5 and ACE2-SHiELD. For each ACE2 evaluation, a three-member initial condition (IC) ensemble of the model (each initialized one day apart) is shown in solid lines, and the reference dataset is shown in dashed lines (e.g., ACE2-ERA5 vs. ERA5 itself). The validation and test periods are shaded in dark gray and light gray, respectively. As a baseline, the ACE-climSST model \citep{WattMeyer2023} forced with the historical SST is also shown for total water path (2-meter air temperature was not predicted by this model). The "forced SST" in a) is the prescribed SST averaged over 45$^{\circ}$S to 45$^{\circ}$N in the SHiELD simulation (ERA5 SSTs are similar though not identical). The R$^2$ of the 81-year series are shown in b) and d). For ACE2-SHiELD, the skill metrics for each of four trained models are shown. Error bars indicate the range over three IC ensemble members for each model. SHiELD reference variability is the R$^2$ computed between the two SHiELD ensemble members.}
    \label{fig:annual_mean_series}
\end{figure}

Figure~\ref{fig:annual_mean_series} shows time series of global- and annual-mean variables for ACE2 and the reference datasets. Both ACE2-ERA5 and ACE2-SHiELD track the long-term trends of their reference datasets closely, which are driven largely by the forced SST trends. Differences in 2-meter air temperature between ERA5 and SHiELD themselves, despite the same SSTs, are largely from disagreement over high-elevation land and polar sea and land ice (not shown). Spatial patterns of long-term trends in the reference dataset are well-matched by ACE2-SHiELD (Appendix \ref{appendix:temp_trend_maps}). Shorter term inter-annual variability of 2-meter air temperature and total water path is also reflected in ACE2's predictions but is slightly muted compared to the reference datasets. The performance of ACE2 is similar between the training and validation periods and the held out test period (shaded light gray). In contrast, the previously trained ACE-climSST \citep{WattMeyer2023} does not reproduce the historical moistening trends (Figure~\ref{fig:annual_mean_series}c) when forced with AMIP SST; it also fails to predict historical warming in other temperature variables that is captured by ACE2 (not shown). 

The ACE2-SHiELD and ACE2-ERA5 models chosen by our checkpoint selection criteria (best inference performance over 1940-2000, see Section~\ref{subsec:training}) have similar skill in predicting inter-annual variability, comparable to the noise floor set by the SHiELD reference variability. Figures~\ref{fig:annual_mean_series}b and \ref{fig:annual_mean_series}d show a scalar skill metric (R$^2$) of the global- and annual mean series, including each of the four models in the training ensemble for ACE2-SHiELD. e.g., ACE2-ERA5 has a mean R$^2$ of 2-meter air temperature of 0.93, while for SHiELD reference variability the R$^2$ is 0.97. However, not all members of the training ensemble for ACE2-SHiELD have the same skill; one of the trained models (labeled "-RS3") has much poorer skill than the other three. 

\subsection{Test period evaluation}

\subsubsection{Climate skill}
\label{subsubsec:holdout_climate_skill}

\begin{figure}[t]
    \centering
    \includegraphics[width=\textwidth]{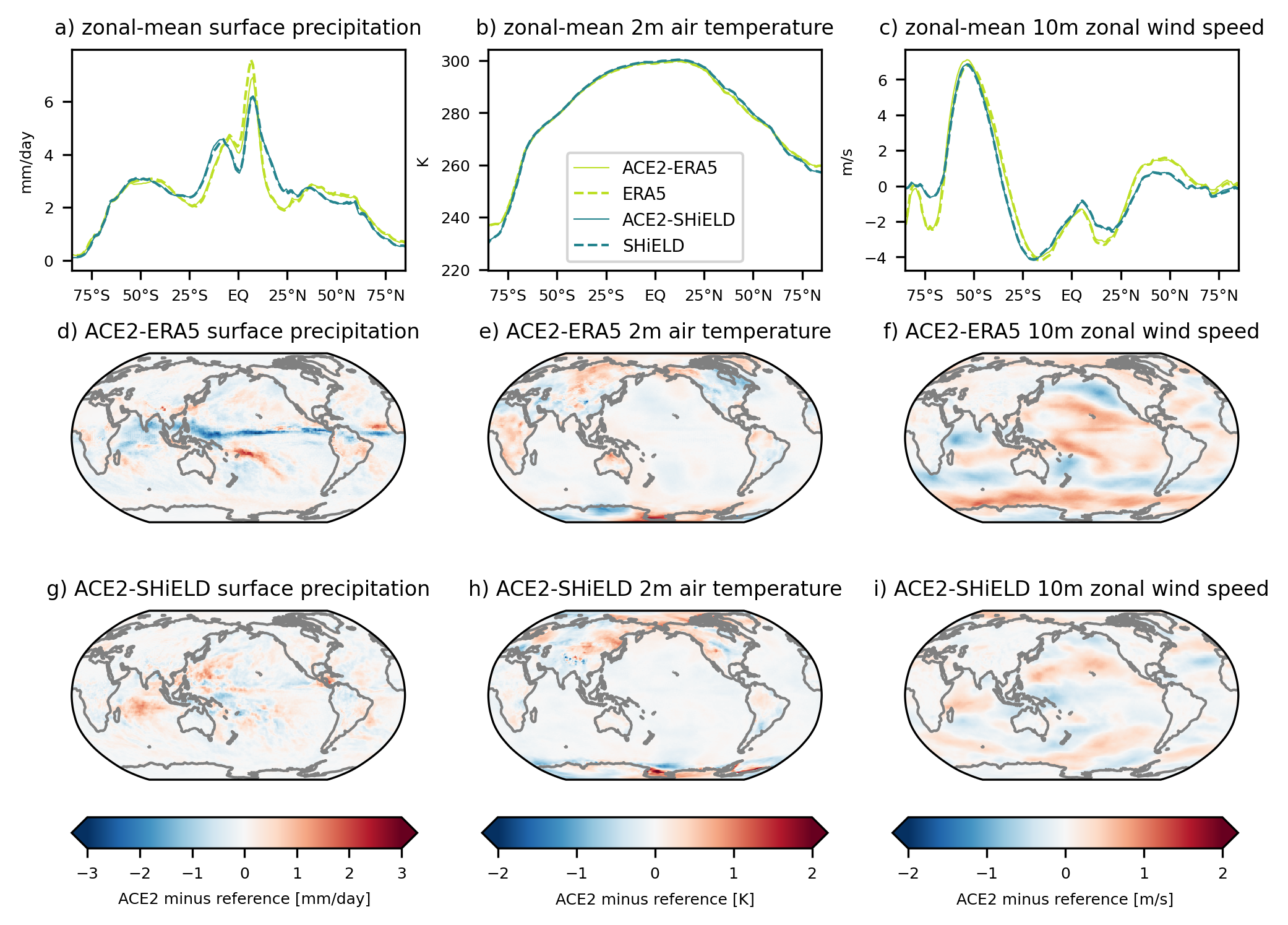}
    \caption{a) - c): Zonal- and time-mean for ACE2 (solid) and its reference datasets (dashed) over test period spanning 2001-01-01 to 2010-12-31, for selected variables. d) - f): ACE2-ERA5 time-mean biases over this time period. g) - i): ACE2-SHiELD time-mean biases over this time period. Results for a single initialization of each ACE2 model are shown.}
    \label{fig:time_zonal_mean}
\end{figure}

We evaluate ACE2's inference performance on a 10-year simulation forced by SSTs and CO$_2$ from the test period 2001-01-01 to 2010-12-31. Figures~\ref{fig:time_zonal_mean}a-c shows the zonal- and time-mean of the ACE2-ERA5 and ACE2-SHiELD predictions. Each model's predictions adhere closely to its reference dataset in zonal- and time-mean, such that ACE2 errors are much smaller in magnitude than the difference between the ERA5 and SHiELD datasets themselves. 

The time-mean bias spatial patterns of ACE2-ERA5 and ACE2-SHiELD are different for surface precipitation and 10-meter wind speed (Figures~\ref{fig:time_zonal_mean}d, g and f, i), but for both models the largest precipitation errors are around the oceanic tropical convergence zones, where time-mean precipitation is large. The models' bias patterns are more similar for 2-meter air temperature (Figures~\ref{fig:time_zonal_mean}e, h) with larger-magnitude temperature biases over high-latitude land and sea ice.
Over ocean regions the temperature biases are smaller, as expected due to their strong coupling with the specified SST.

\begin{figure}[t]
    \centering
    \includegraphics[width=\textwidth]{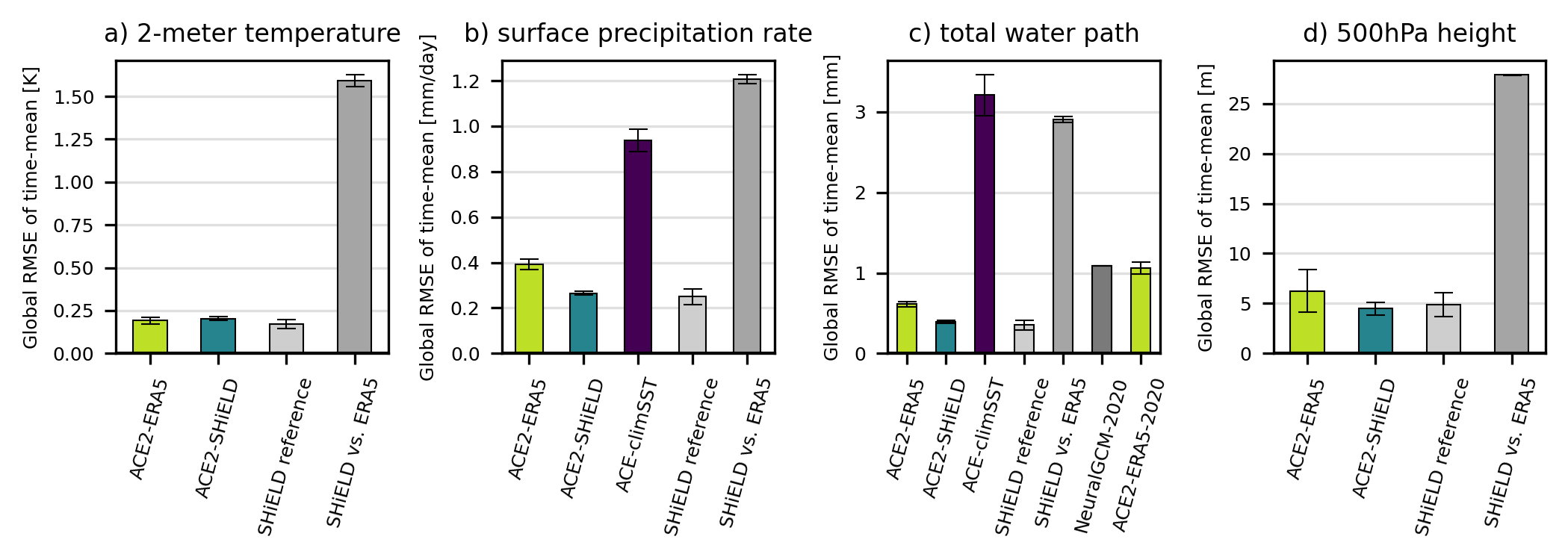}
    \caption{Global RMSE between the time-mean of ACE2 and its reference dataset (ERA5 or SHiELD). Error bars indicate the 95\% confidence interval based on the IC ensemble. Also included are NeuralGCM error against ERA5, SHiELD reference variability, the error of ACE-climSST evaluated against the SHiELD dataset, and the error of the SHiELD simulations against ERA5. ACE-climSST did not predict 2-meter temperature or 500hPa height. NeuralGCM \citep{Kochkov2024} results are only available for total water path for a single year (2020), and so we also show 2020-only results of ACE2-ERA5.}
    \label{fig:time_mean_RMSE_10yr}
\end{figure}

To quantify the magnitudes of the biases above, global time-mean RMSEs (Equation~\ref{eq:time_mean_RMSE}) of key surface fields over the 10-year test period are shown in Figure~\ref{fig:time_mean_RMSE_10yr}. The errors of ACE2-ERA5 are computed with respect to the ERA5 dataset, while the ACE2-SHiELD and ACE-climSST errors are computed with respect to SHiELD. For all variables, the ACE2 models easily outperform the prior ACE model (ACE-climSST; \cite{WattMeyer2024}) and their errors are much smaller than the difference between the SHiELD and ERA5 datasets. To enable comparison with NeuralGCM, for which the time-mean error of total water path over a 1-year simulation was reported (c.f. Figure 4i of \cite{Kochkov2024}) we run an analogous ACE2-ERA5 simulation spanning 2020, a period not used for training or validation. ACE2-ERA5 has similar error as NeuralGCM, about 1.05mm versus 1.09 mm, respectively, over this period (Figure~\ref{fig:time_mean_RMSE_10yr}c). 

The error magnitudes of ACE2-ERA5 and ACE2-SHiELD against their reference datasets are similar; the model with the smaller error depends on the variable. In addition, the error magnitudes for ACE2-SHiELD are typically only 1.1-1.5 times the SHiELD reference variability (which is the magnitude of differences between the two SHiELD ensemble members, sampled over different 10-year periods). That is, by this metric, the 10-year mean climate of ACE2 is nearly indistinguishable from that of the reference model.

The ACE2-SHiELD training ensemble shows non-trivial variability between models; the selected model ("ACE2-SHiELD") slightly outperforms the other models ("ACE2-SHiELD-RS0", "-RS1", "-RS3") over the test period (Figure~\ref{fig:time_mean_RMSE_with_ensemble}). See Appendix~\ref{appendix:discussion_seed_variability} on model selection for more information.

\subsubsection{Atmospheric response to ENSO variability}

\begin{figure}[t]
    \centering
    \includegraphics[width=\textwidth]{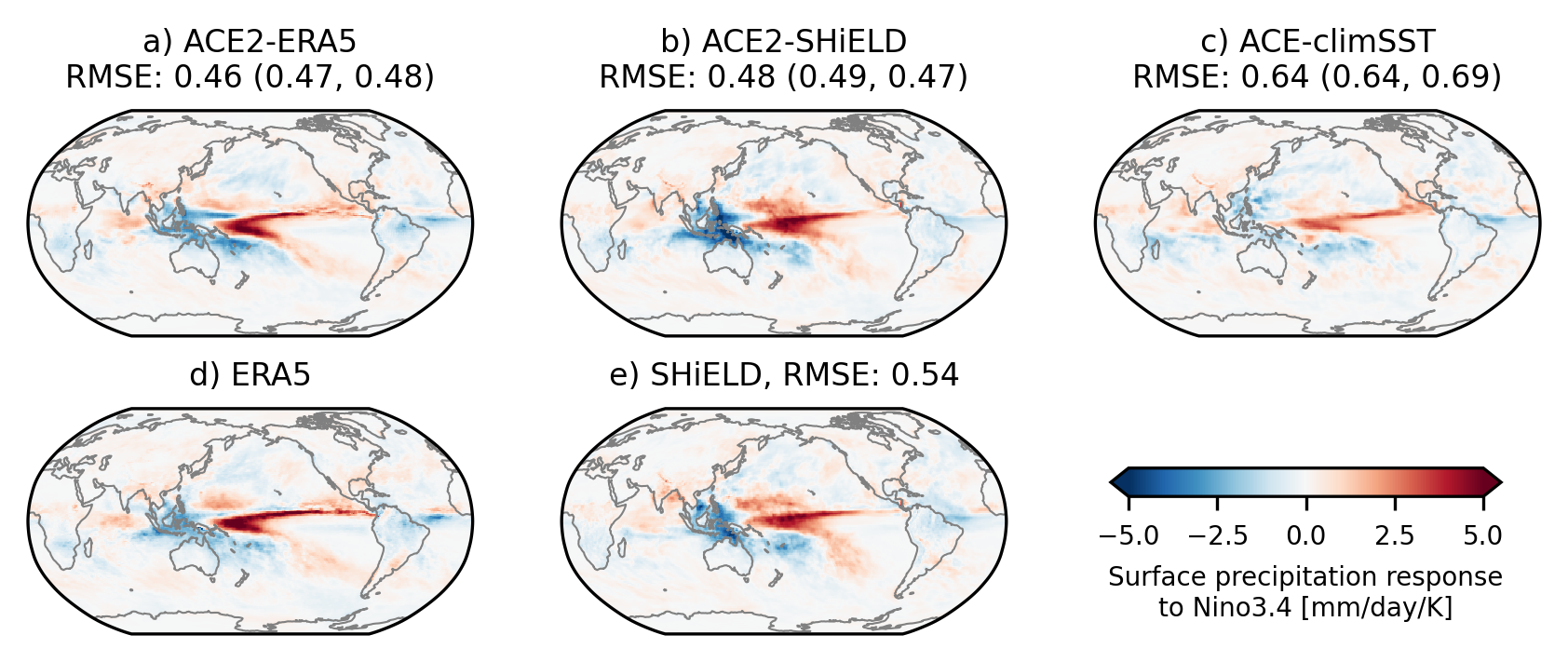}
    \caption{Maps of regression coefficients of predicted and reference dataset surface precipitation against the Ni\~no 3.4 index over the 10-year test period. Single model initializations are shown. a) ACE2-ERA5, b) ACE2-SHiELD, c) ACE-climSST evaluated on SHiELD, d) ERA5 reference, e) SHiELD reference, all for the 10-year test period. Titles of panels a-c indicate the RMSE of the predicted map against its reference map; the numbers in parenthesis are for the two other initializations that are not shown. For e), the SHiELD reference variability is calculated as the RMSE between the regression coefficient maps of the two ensemble members.}
    \label{fig:enso_precip}
\end{figure}

We compute the atmospheric response to the El Ni\~no-Southern Oscillation (ENSO; \cite{Trenberth1997}) by regressing the predicted variables onto the Ni\~no 3.4 index (see Eq. \ref{eq:nino_coefficient}). Maps of the ENSO-regressed surface precipitation rate for ACE2-ERA5, ACE2-SHiELD, their reference datasets, and ACE-climSST are shown for the 10-year test period. ACE2 reliably reproduces the canonical response of surface precipitation to Ni\~no 3.4 variability \citep{Trenberth1997} in which positive Ni\~no 3.4 is associated with increased precipitation in the central tropical Pacific and western Indian Ocean, and decreased precipitation over the  maritime continent and tropical Atlantic (Figure~\ref{fig:enso_precip}).  Furthermore, ACE2 clearly reproduces the details of the Nino3.4 regression maps in the respective ERA5 and SHiELD reference datasets.

In contrast, the previous ACE-climSST predictions show a somewhat skillful but muted precipitation response to Ni\~no 3.4 when evaluated using SHiELD forcing (Figure~\ref{fig:enso_precip}c), demonstrating the value of ACE2 over ACE-climSST, most notably due to training on datasets with historical SST variability. The RMSE of the precipitation ENSO regression maps for ACE2-ERA5 (0.46 mm/day/K) and ACE2-SHiELD (0.48 mm/day/K) are smaller than that of ACE-climSST (mean 0.64 mm/day/K), and are comparable to the internal variability of this regression map in SHiELD (0.54 mm/day/K). A similar result is found for outgoing longwave radiation at top of atmosphere (Figure~\ref{fig:enso_olr}).

Maps of ENSO coefficients for ACE2 rollouts spanning the entire 81-year ERA5/SHiELD period (not shown) are qualitatively similar to those for the 10-year test period, showing that the learned response to ENSO is robust.

\subsubsection{Tropical cyclone climatology}
\label{subsec:tropical-cyclones}
Tropical cyclones are particularly damaging weather phenomena whose characteristics, such as strength and intensification rate, are projected to change with global warming \citep{Elsner2008, Bhatia2019, Vecchi2021}. Their accurate representation would be a valuable feature of climate model emulators to allow the assessment of changes in these properties as a function of changing boundary conditions. In this section, we compare the strength, frequency, and location of tropical cyclone-like features in the ERA5 dataset, the ACE2-ERA5 emulator and, for comparison, the C96 (approximately 100$\,$km resolution) SHiELD atmospheric model. However we note the SHiELD atmospheric model at C96 resolution is not expressly designed or intended to accurately represent tropical cyclones.

The features are detected using 1° horizontal resolution data, although tropical cyclones are not well resolved at this horizontal resolution (e.g. their strength is often underestimated \citep{Hodges2017} or they may be simply not detected). We use the TempestExtremes package and apply the default setting recommended for detecting tropical cyclones (Section 3.2 of \cite{Ullrich2021}), noting that these defaults were originally tuned for ERA5 at 0.25° resolution. One exception is that instead of using upper-level geopotential thickness (Z300 minus Z500) to detect warm cores aloft, we use upper tropospheric temperature since ACE2 does not directly predict geopotential height. Specifically, we use $T_3$, which is the mean temperature between about 250$\,$hPa and 400$\,$hPa (see Tables~\ref{table:variables} and \ref{table:verticalcoord}). Instead of requiring a thickness decrease away from the tropical cyclone center, we require a temperature decrease of 0.4$\,$K. Assuming hydrostatic balance, this is approximately equal to the 58.8$\,$m$^2$s$^{-2}$ thickness decrease suggested in \cite{Ullrich2021}.

\begin{figure}[t]
    \centering
    \includegraphics[width=\textwidth]{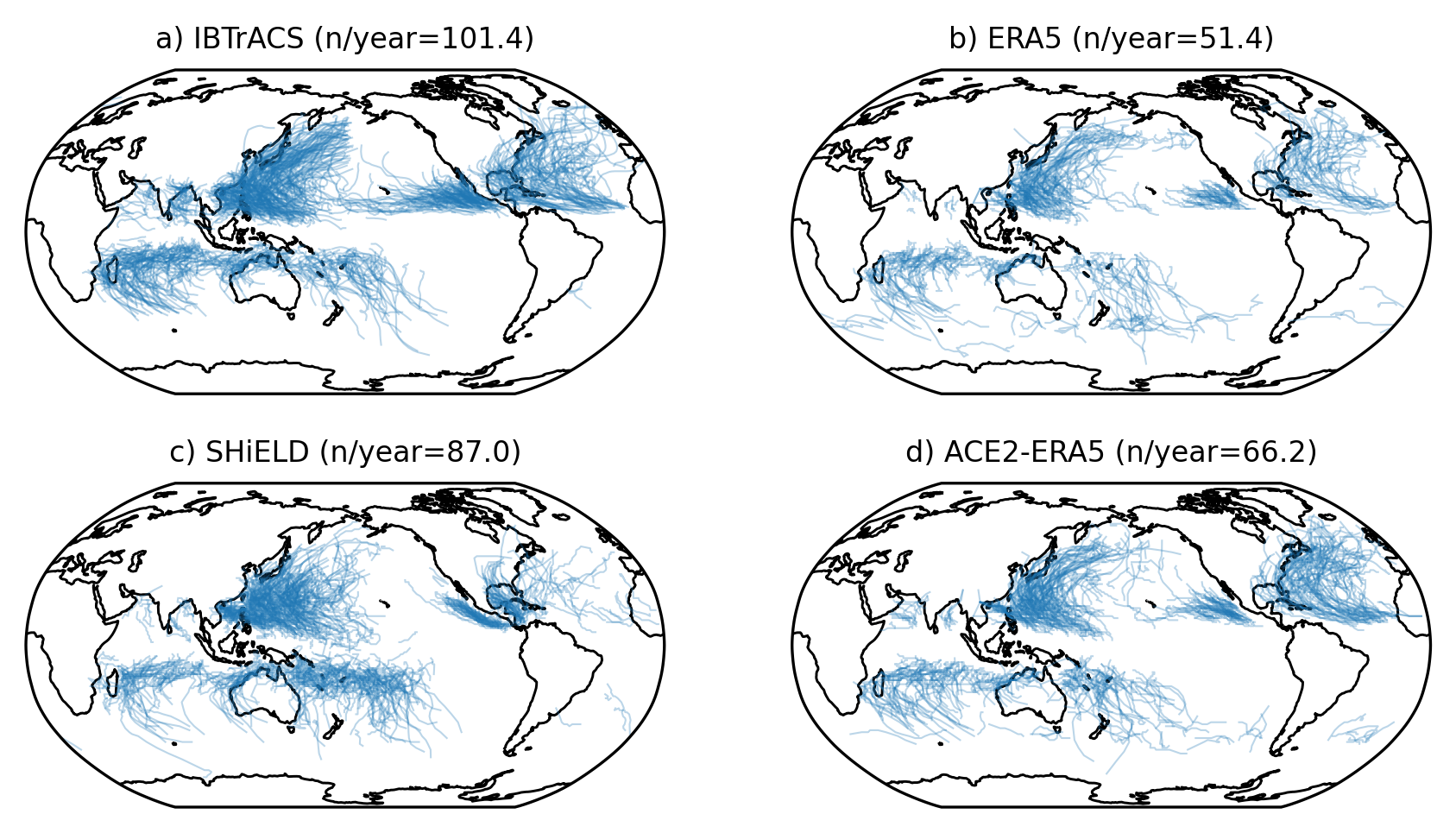}
    \caption{Tracks of tropical cyclone-like features over the 2001-2010 period for a) the IBTrACS dataset, b) ERA5, c) the C96 SHiELD model and d) ACE2-ERA5. The tracks for b)-d) are determined based on minima in sea-level pressure along with maxima in upper troposphere temperature. See main text for details. The average number of tropical cyclones across the globe per year is shown in the title of each panel, although the IBTrACS dataset is not directly comparable to the detections in other panels which use a tracking algorithm applied to 1° resolution data.}
    \label{fig:tc-tracks}
\end{figure}

Figure~\ref{fig:tc-tracks} shows the tropical cyclone tracks detected for ERA5, ACE2-ERA5 and SHiELD according to the above criteria as well as those in the IBTrACS database \citep{Knapp2010,IBTRACS} for the 10-year test period (2001-2010). The number of cyclones detected per year globally is shown in the title of each panel, although we note that this quantity is sensitive to the parameters chosen for the detection algorithm used in Figures~\ref{fig:tc-tracks}b-d. That said, since the same detection parameters are used for ERA5, ACE2-ERA5 and SHiELD, we can compare this quantity between these datasets. Globally, ACE2-ERA5 overpredicts tropical cyclone frequency by about 28\% compared to its target dataset ERA5. The SHiELD atmospheric model predicts about 69\% more tropical cyclones that ERA5 at the given resolution, which may be more in line with the true frequency of tropical cyclones (Figure~\ref{fig:tc-tracks}a). Regionally, ACE2-ERA5 closely matches the basin-by-basin frequency of tropical cyclones in the ERA5 dataset (Figure~\ref{fig:tc-tracks}). Compared to ERA5 and IBTrACS, the SHiELD atmospheric model has too few tropical cyclones in the North Atlantic a difference possibly caused by a biased mean circulation in the model. Overall, this analysis suggests the ACE2-ERA5 emulator accurately captures the regional frequency of tropical cyclone-like events in the ERA5 dataset.

A possible concern with our evaluation framework is that ACE2-ERA5 is forced with observed sea surface temperatures that contain a signature of past tropical cyclones, which can leave behind a cold wake \citep{Price1981}. Hypothetically, the machine learning emulator could learn to generate tropical cyclones based on the prescribed sea surface temperature signature. However, when we force ACE2-ERA5 with a climatological sea surface temperature dataset, we recover a very similar frequency and distribution of tropical cyclones as when we force it with historical sea surface temperature, showing that this is not the case (Figure~\ref{fig:tc-climsst-tracks}).

The strength of the detected tropical cyclones, as measured by minimum sea level pressure and maximum 10$\,$m wind speed, is also accurately emulated by ACE2-ERA5 (Fig.~\ref{fig:tc-statistics}) when compared to the ERA5 dataset. The SHiELD model tends to produce more cyclones with strong (>30$\,$m/s) near-surface wind speeds.

\subsubsection{Tropical precipitation variability}
Prior work has confirmed that ACE is able to closely replicate the precipitation variability in a coarse resolution atmospheric model \citep{Duncan2024}. Here we show a brief analysis of tropical precipitation variability focused on ACE2-ERA5 since the ERA5 dataset contains variability, such as equatorial Kelvin waves or the Madden-Julian Oscillation, which is often missing or too weak in coarse resolution atmospheric models (c.f. Fig 17d of \cite{Golaz2022}, \cite{Ahn2020}).

\begin{figure}[t]
    \centering
    \includegraphics[width=\textwidth]{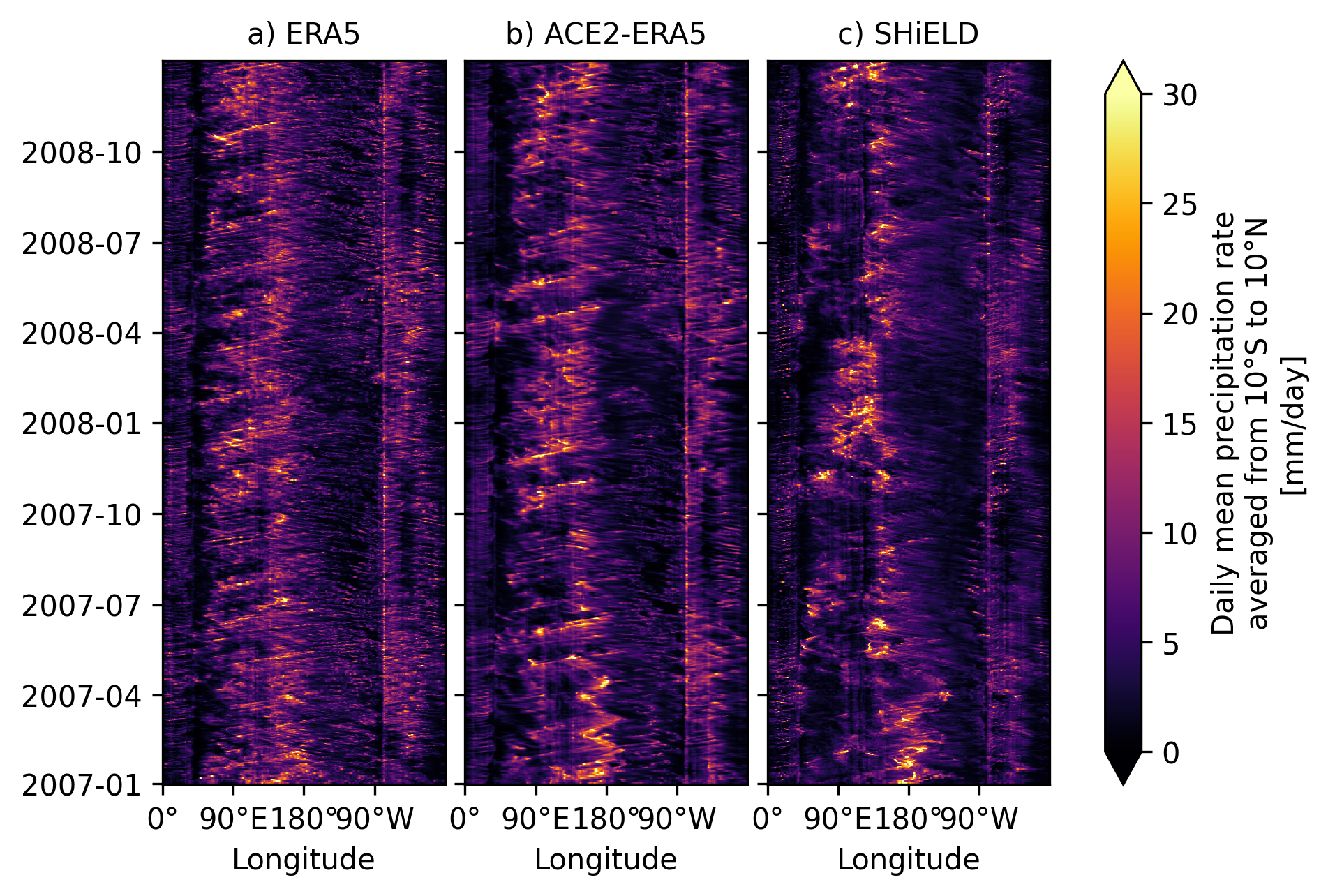}
    \caption{Daily-mean precipitation rate averaged between 10°S and 10°N over the 2007-2008 period for (a) ERA5 and (b) the 10-year ACE2-ERA5 run initialized on 2001-01-01 and (c) the first ensemble member of the SHiELD AMIP simulation.}
    \label{fig:daily-precip}
\end{figure}

Figure~\ref{fig:daily-precip} shows the tropical-mean precipitation over longitude and time for the 2007-2008 period, which contained several strong Madden Julian Oscillation (MJO) events in the observed record \citep[e.g.][]{Hagos2011} that are apparent in the ERA5 dataset, for example during December 2007 (Fig.~\ref{fig:daily-precip}a). The shown ACE2-ERA5 and SHiELD simulations (Figs.~\ref{fig:daily-precip}b-c) are initialized in 2001 and 1939 respectively, so we do not expect the timing of events to coincide between the three datasets. However, it is notable that the spatio-temporal variability of the ERA5 dataset is much more closely captured by ACE2-ERA5 than it is by SHiELD. For example, relatively small-scale eastward propagating Kelvin waves \citep{Wheeler1999} exist in both the ERA5 and ACE2-ERA5 precipitation variability, but are less apparent in SHiELD. ACE2-ERA5 does show some notable differences from ERA5, for example generally being smoother in longitude and time.

\begin{figure}[t]
    \centering
    \includegraphics[width=\textwidth]{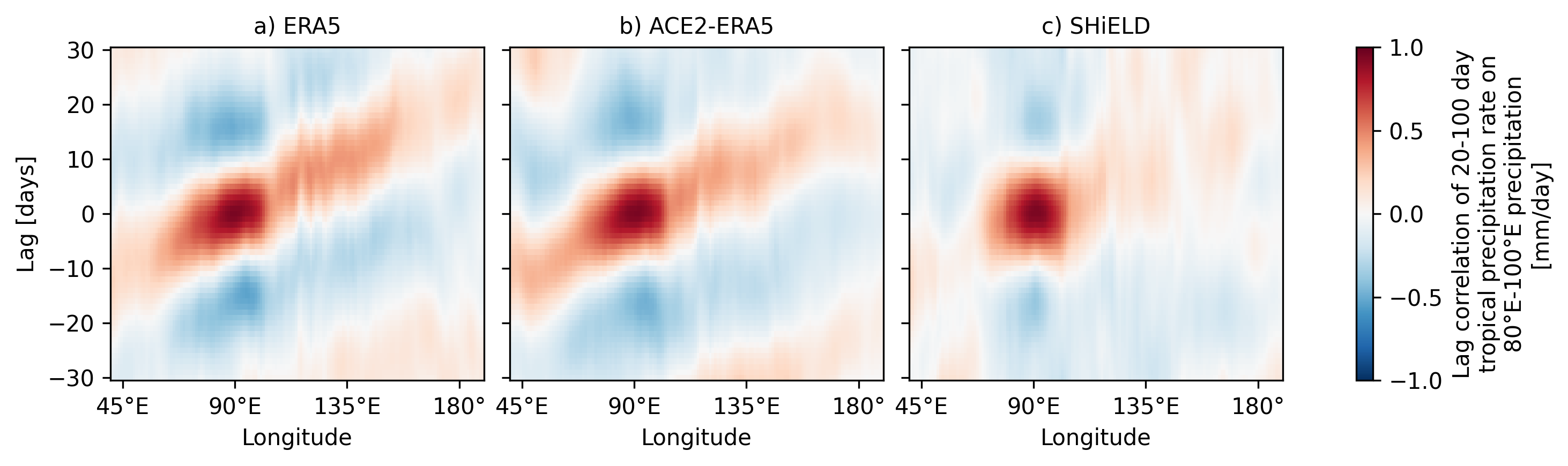}
    \caption{Lag correlation of $P_{10^{\circ}S-10^{\circ}N}^{20-100\mathrm{ day}}$ at all longitudes with $P_{10^{\circ}S-10^{\circ}N}^{20-100\mathrm{ day}}$ averaged from 80°E to 100°E \citep[e.g.][]{Waliser2009}.  $P_{10^{\circ}S-10^{\circ}N}^{20-100\mathrm{ day}}$ is the surface precipitation rate averaged from 10°S to 10°N, and then filtered with a 20-100 day bandpass filter. Calculated over the 2001-2010 period from (a) ERA5, (b) the test period run of ACE2-ERA5 and (c) the first ensemble member of the SHiELD AMIP run.}
    \label{fig:mjo}
\end{figure}

To more explicitly compare the representation of the MJO, the dominant mode of intraseasonal variability in the tropics \citep{Zhang2005}, we compute a lag-correlation diagnostic which demonstrates the eastward movement of precipitation on the MJO timescale (20-100 days) around the Indian Ocean and Maritime Continent \citep{Waliser2009}. Specifically, we first compute $P_{10^{\circ}S-10^{\circ}N}^{20-100\mathrm{ day}}$, which is the surface precipitation rate averaged between 10°S and 10°N and bandpass filtered between 20 and 100-day variability. We then compute the lag correlation of $P_{10^{\circ}S-10^{\circ}N}^{20-100\mathrm{ day}}$ at all longitudes with that over the western Indian Ocean (80°E and 100°E). Figure~\ref{fig:mjo} shows this lag-correlation for ERA5, ACE2-ERA5 and SHiELD, in all cases computed over 2001-2010. This demonstrates the eastward propagation of of the MJO in ERA5 (Fig.~\ref{fig:mjo}a) while the SHiELD model lacks coherent eastward propagation of precipitation variabilty in this region (Fig.~\ref{fig:mjo}c), a fairly common and longstanding issue of coarse resolution global atmospheric models \citep{Waliser2003,Kim2009,Ahn2020}. However, ACE2-ERA5 shows an eastward propagation of the MJO consistent with ERA5, both in terms of phase speed and longitudinal extent. This lends further credibility to the realism of ACE2-ERA5 emulator's representation of tropical variability on subseasonal timescales.

\subsubsection{Polar stratospheric variability}
Existing machine learning models for weather prediction either do not explicitly resolve the stratosphere \citep{Weyn2020,Karlbauer2024}, do not report on skill in the stratosphere \citep{Bi2023,Chen2023}, or show relatively worse short-term predictive performance in the stratosphere compared to lower vertical levels \citep{Lam2023}. The uppermost vertical layer of ACE2 represents a mass-weighted integral of atmospheric properties (temperature, horizontal winds and moisture) between approximately 50 hPa and the top of atmosphere (Table~\ref{table:verticalcoord}). Therefore, we are able to evaluate the representation of large-scale stratospheric processes. In this section, we focus on comparing polar stratospheric variability in ERA5 and ACE2-ERA5. The variability in the strength of the stratospheric polar vortex---as measured by the zonal mean wind $u_0$ in ACE2's top vertical layer at 60° latitude---is the dominant mode of sub-seasonal variability in the stratosphere. It is an important source of sub-seasonal to seasonal predictability \citep{Baldwin2001} and is a strong control on ozone chemistry, resulting in the ozone hole being most evident in the Southern Hemisphere \citep{Solomon1999}.

\begin{figure}[t]
    \centering
    \includegraphics[width=0.9\textwidth]{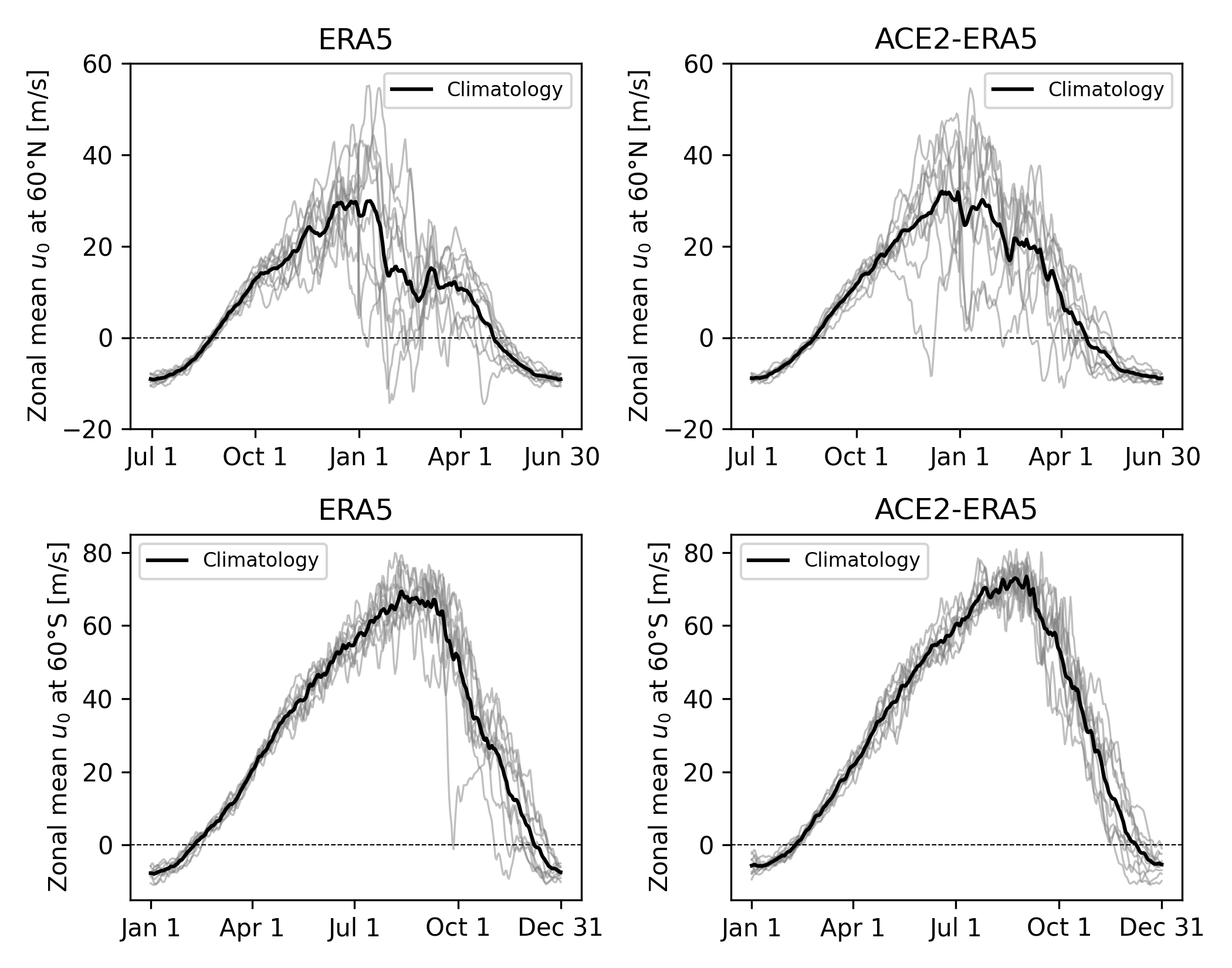}
    \caption{Annual cycle of zonal-mean $u_0$ (eastward wind vertically integrated from $\sim$50hPa to top of atmosphere) at (top row) $60^{\circ}$N and (bottom row) $60^{\circ}$S for (left column) ERA5 and (right column) ACE2-ERA5. For ERA5, each of the years from 2001-2010 test period are plotted. For ACE2-ERA5, a simulation is initialized from ERA5 on 2001-01-01 and run for 10 years. Individual grey lines show each year, while the bold black line shows the average over the 10-year period.}
    \label{fig:polar-stratosphere}
\end{figure}

ACE2-ERA5 reproduces the expected seasonal asymmetry in mean polar stratospheric vortex strength and variability (Figure~\ref{fig:polar-stratosphere}). By overlaying the zonal mean $u_0$ at 60°N and 60°S for each of the 10 years from the test period, we see the expected variability in the Northern Hemisphere exists in ACE2-ERA5. This includes sudden stratospheric warming events in which the strength of the vortex rapidly decreases and the zonal-mean flow reverses. As expected, in the Southern Hemisphere, the average winds are stronger while also being less variable from year to year. With only ten years for comparison, it is difficult to quantitatively compare the statistics of variability between ERA5 and ACE2-ERA5, but the qualitative behavior shown in Figure~\ref{fig:polar-stratosphere} is promising. Longer simulations, which overlap with the training and validation periods, demonstrate good agreement between the 5th and 95th percentiles of $u_0$ at 60°s and 60°N (not shown).

While ACE2-ERA5 shows some variability of near-equatorial stratospheric winds between eastward and westward with approximately the same magnitude as the observed quasi-biennial oscillation \citep{Anstey2022}, the variability is irregular and does not have the correct period (not shown).

\subsubsection{Weather skill}
\begin{figure}[t]
    \centering
    \includegraphics[width=\textwidth]{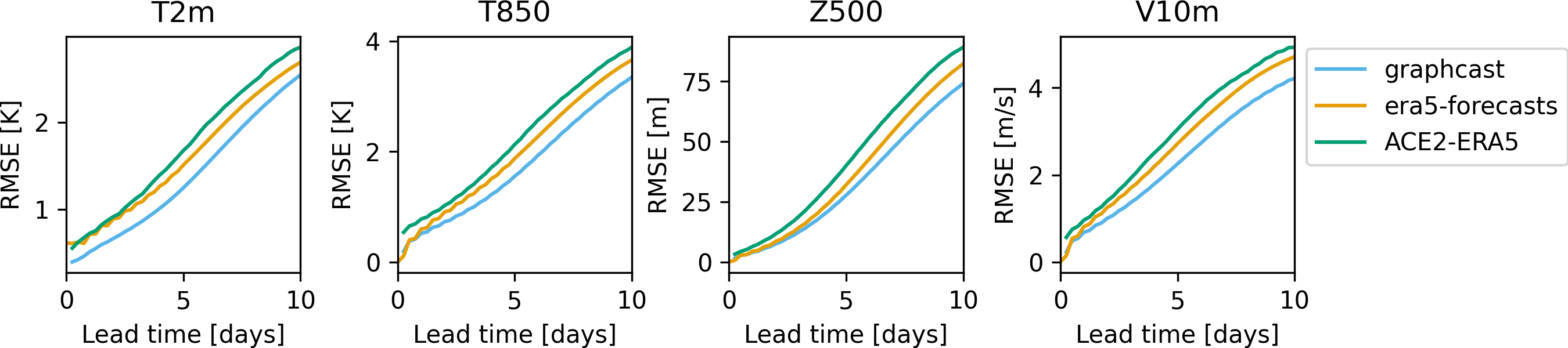}
    \caption{RMSE of ACE2-ERA5 during 2020, compared to GraphCast and IFS initialized from ERA5 (``era5-forecasts''). In order to make a fair comparison, for these ACE2-ERA5 simulations the sea surface temperature and surface type fractions are kept fixed at their initial values instead of being prescribed throughout the simulation.}
    \label{fig:weather-skill}
\end{figure}

Although accurate weather forecast skill was not a primary objective of this work, in this section we assess ACE2-ERA5's medium range global forecast skill. Figure~\ref{fig:weather-skill} shows global RMSE averaged over 10-day forecasts initialized throughout the 2020 period for $T_{2m}$, $T_{850}$, $Z_{500}$ and $v_{10m}$ (see Table~\ref{table:variables} for definitions). As baselines, we use Graphcast \citep{Lam2023} and the ``ERA5 forecasts'' as provided by WeatherBench 2.0 \citep{Rasp2024}, both compared with the ERA5 dataset. The ``ERA5 forecasts'' are forecasts using ECMWF's IFS model, with the same model version used to produce the ERA5 reanalysis and initialized from ERA5 snapshots to provide a more direct comparison with models such as ACE2-ERA5. The ACE2-ERA5 forecasts in Fig.~\ref{fig:weather-skill} correspond to 48 initializations equally spaced across 2020, while Graphcast and era5-forecasts consist of forecasts initialized at 0Z and 12Z on every day of 2020. Figure~\ref{fig:weather-skill} shows that ACE2-ERA5 is slightly behind the ERA5-version of IFS (e.g. half a day at 5-day lead time for 850$\,$ temperature) and further behind Graphcast by another day. Although there are a variety of differences between ACE2-ERA5 and Graphcast, we believe the most substantial to be 1) the different vertical coordinate (ACE2-ERA5 uses a terrain-following coordinate, Graphcast a pressure coordinate) and 2) the architecture underlying each model (SFNO and Graph Neural Network respectively). Preliminary work has found that using a different architecture but keeping ACE2's terrain-following vertical coordinate can lead to a significant increase in weather forecast skill, but not mean climate skill (not shown), suggesting that the architecture difference is likely the more important one.

\subsubsection{Millennial timescale stability}
\begin{figure}[t]
    \centering
    \includegraphics[width=\textwidth]{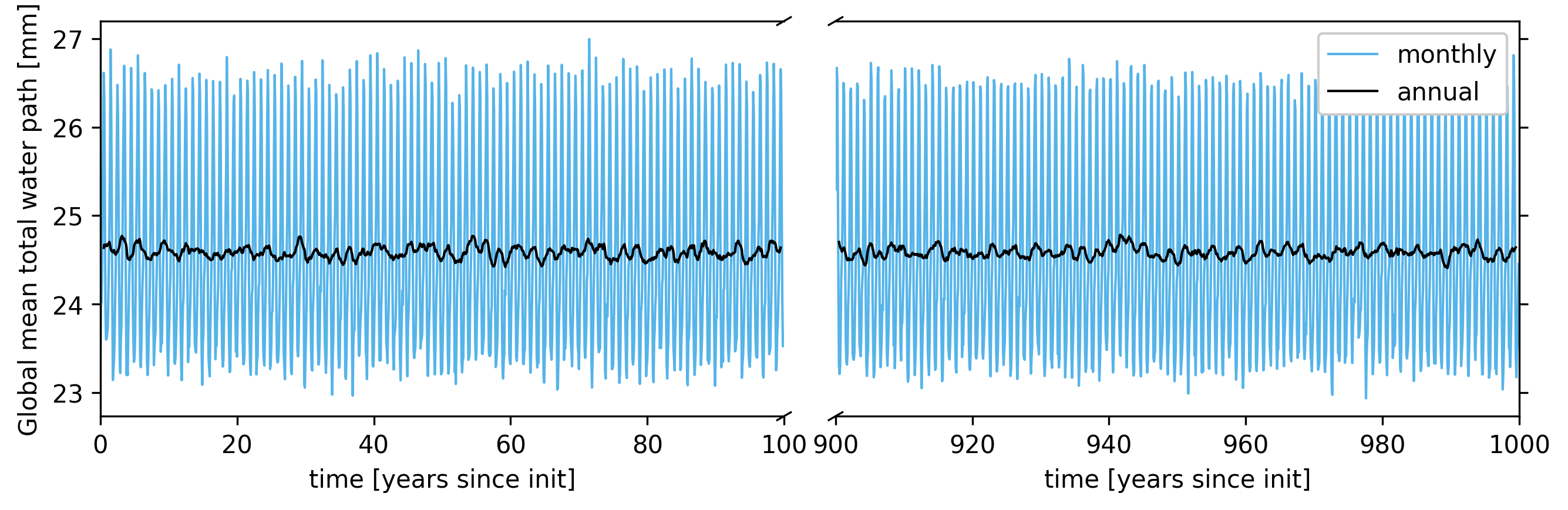}
    \caption{The global mean total water path for the first and last 100 years of a 1000-year long simulation with ACE2-ERA5 forced with 1990-2020 climatological mean sea surface temperatures, land type fractions and CO$_2$. Shown for (blue) monthly mean and (black) annual mean.}
    \label{fig:1000yr-run}
\end{figure}

To test the stability of ACE2 models over a longer duration than the length of the AMIP forcing dataset (about 80 years), we compute a climatological forcing dataset which can be repeated indefinitely. This is computed by averaging surface temperature, CO$_2$ and the surface type fractions from ERA5 over the 1990-2020 period, resulting in a 6-hourly climatology estimate. We then initialize an ACE2-ERA5 simulation from ERA5 on 2001-01-01 and run it for 1000 years, forced by the annually repeating climatological dataset. No signs of instability (i.e. indefinitely growing errors) are seen in this 1000-year run, and the time-mean climate is nearly identical between different 100-year periods of the simulations. As an example, Figure~\ref{fig:1000yr-run}, shows the global-mean total water path timeseries for the first 100 and last 100 years of the simulation. There is no long term drift in total atmospheric moisture, and the seasonal cycle remains of consistent amplitude throughout the simulation. This is a noteworthy improvement on ACE-climSST, where 100-year simulations showed unrealistic fluctuations in the amplitude of the global-mean seasonal cycle \cite[c.f. Figure 10 of][]{WattMeyer2023}.

\subsection{Learning at coarser horizontal resolution}
\label{sec:4-degree}

Traditional climate models often achieve improved skill at increasing resolution, as physical processes are more accurately represented. However, this is not necessarily the case for coarse emulators of a climate model without an explicit representation of atmospheric processes. 
Here we compare the performance of ACE2 trained on the SHIELD AMIP dataset coarsened to 4-degree resolution against the coarsened output of ACE2 trained at 1-degree resolution (as presented in Section \ref{sec:results}). Ideally, the climate of the 4-degree ACE2 emulator could be just as skillful as that of the 1-degree emulator, but is this achievable in practice?


\begin{figure}[t]
    \centering
    \includegraphics[width=\textwidth]{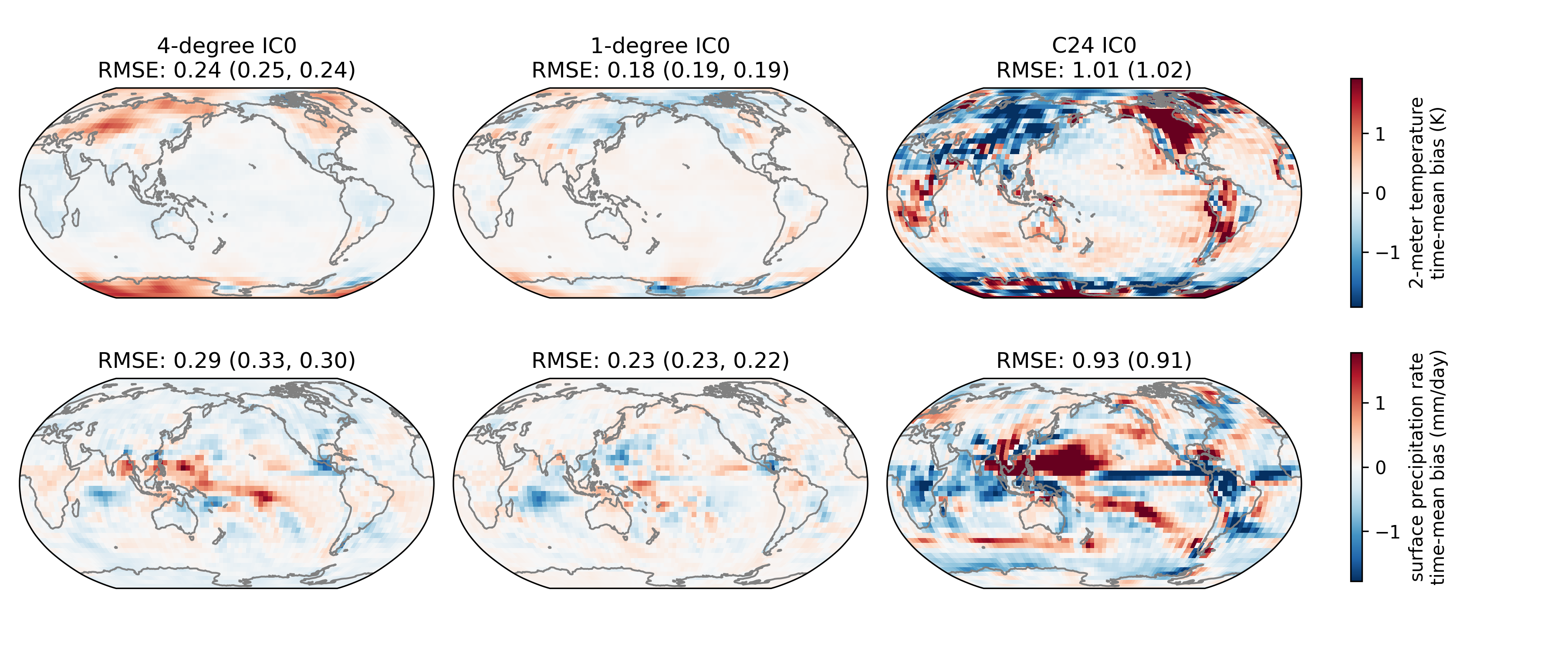}
    \caption{Single initial condition time-mean biases of $T_{2m}$ and precipitation for 10-year inference using 1$^\circ$ and 4$^\circ$ ACE2-SHiELD models and the C24 (4$^\circ$) SHiELD baseline model, with respect to the C96 (1$^\circ$) SHiELD model. 1$^\circ$ values are area-weighted block-coarsened by a factor of 4 prior to computing RMSE. Values are shown for the same time period and ensemble configuration as in Figure \ref{fig:enso_precip}. RMSE is shown for the ensemble member shown in the map, with values for the other two members shown in parentheses.}
    \label{fig:amip-4deg-bias-map}
\end{figure}

\begin{figure}[t]
    \centering
    \includegraphics[width=\textwidth]{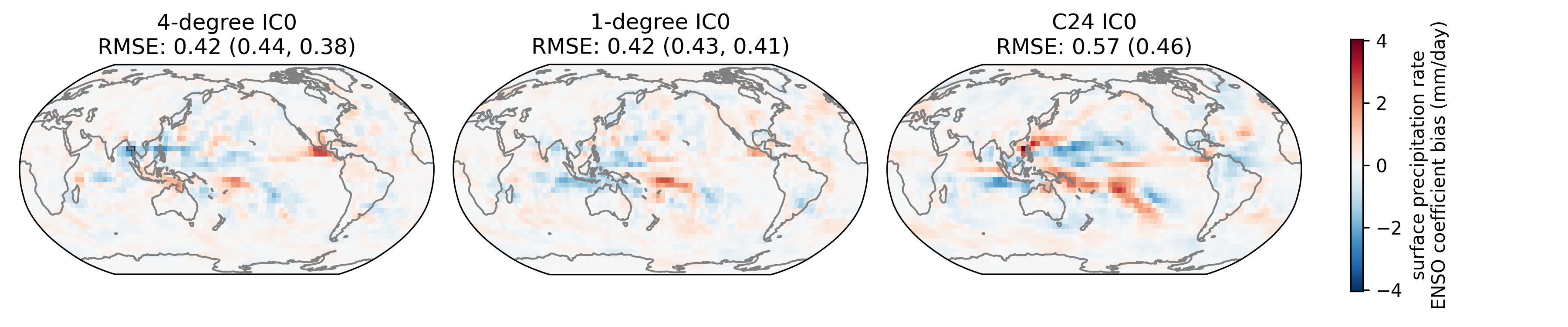}
    \caption{Bias of single initial condition ENSO regression coefficient maps of surface precipitation rate for 3-member 10-year inference using 1 and 4-degree ACE2 models and a C24 (4$^\circ$) SHiELD baseline model, with respect to the C96 SHiELD model. 1-degree values are area-weighted block-coarsened by a factor of 4. Values are shown for the same time period and ensemble configuration as in Figure \ref{fig:enso_precip}. RMSE is shown for the ensemble member shown in the map, with values for the other two members shown in parentheses.}
    \label{fig:amip-4deg-enso-map}
\end{figure}

With identical training and inference regimes, the time-mean ensemble-mean biases of 2~m temperature ($T_{2m}$) and precipitation have slightly higher magnitudes for the models trained at 4$^\circ$ resolution compared with 1$^\circ$ resolution (Figure \ref{fig:amip-4deg-bias-map}). Both have biases much smaller than a C24 (approximately 400$\,$km resolution) SHiELD baseline simulation. The largest $T_{2m}$ biases are at high latitudes. $T_{2m}$ biases over open ocean regions are minimal, as physically expected due to strong coupling of $T_{2m}$ with SST. The biases in time-mean precipitation are largest at low latitudes, in regions of large mean precipitation. We use the C24 SHiELD as a baseline because coarsening spatial resolution is a common strategy to decrease the computational cost of physics-based atmospheric models. However, ACE2 is still about 25x more energy efficient than C24 SHiELD and it is about 700x more energy efficient than C96 SHiELD (Table~\ref{table:inferencecost}).

The patterns of precipitation variability regressed on ENSO variability have similar RMSE amplitudes for the 4$^\circ$ emulator and the 1$^\circ$ emulator, and their biases with respect to the C96 SHiELD model share many of the same spatial structures (Figure \ref{fig:amip-4deg-enso-map}). Both have lower biases with respect to the C96 SHiELD model than the C24 SHiELD baseline.

This similarity in skill between ACE2 trained at 1$^\circ$ and 4$^\circ$ is encouraging because it suggests that, unlike for physics-based climate models, a computationally light coarse emulator that might be attractive for paleoclimate or marine biogeochemistry applications can simulate coarse-scale climate features almost as well as a more expensive, memory-intensive fine-grid emulator.  

\begin{figure}[t]
    \centering
    \includegraphics[width=\textwidth]{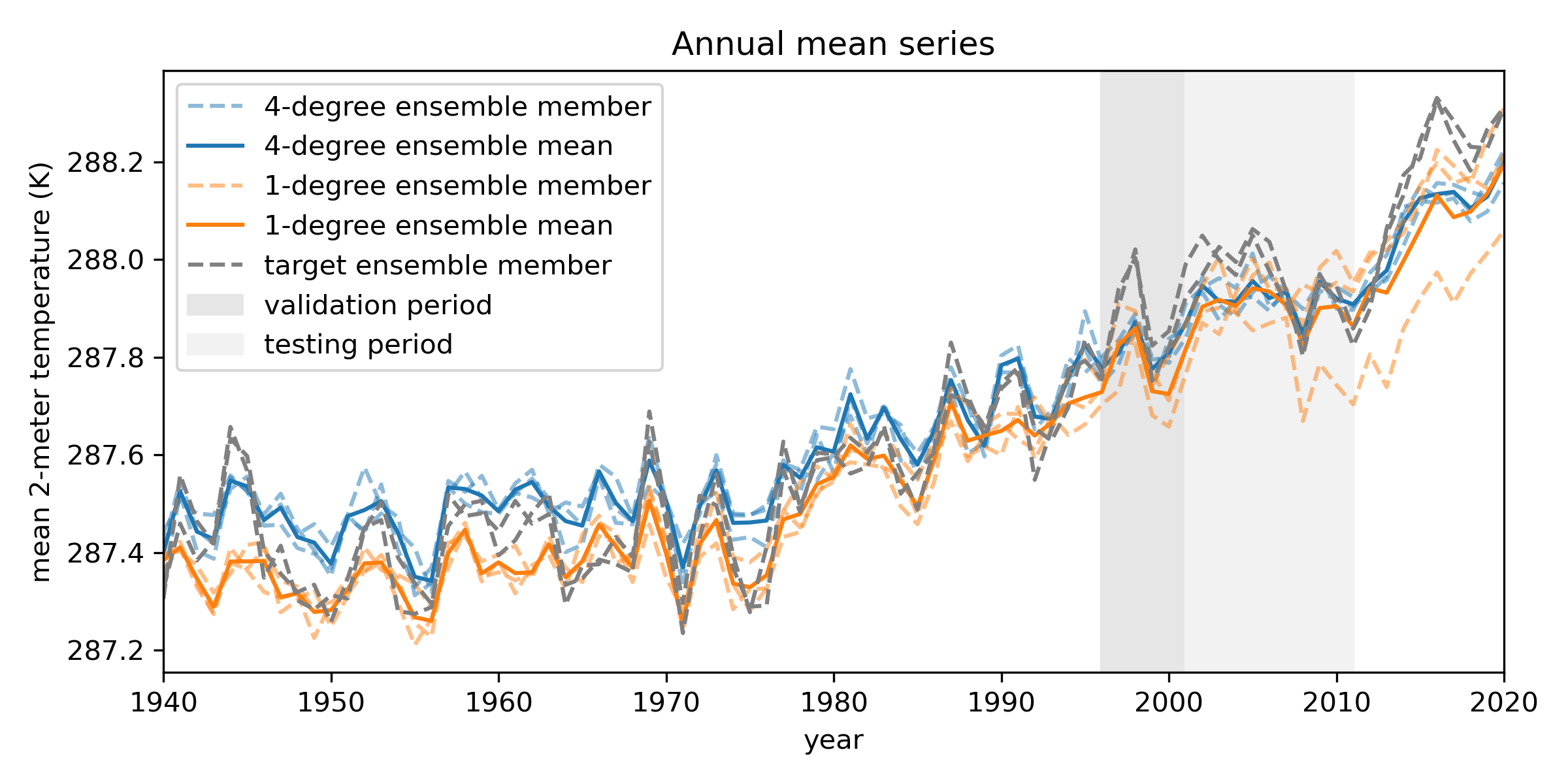}
    \caption{Annual and global mean 2-meter temperature for 81-year inference using 3-member initial condition ensembles of 1 and 4-degree ACE2-SHiELD models. Values are shown for the same time period and ensemble configuration as shown in Figure \ref{fig:annual_mean_series}.}
    \label{fig:amip-4deg-annual-mean}
\end{figure}

The 1$^\circ$ and 4$^\circ$ models show a similar ability to reproduce the long-term trend and interannual variability (Figure \ref{fig:amip-4deg-annual-mean}). Both models show reduced interannual variability over shorter timescales. Over the 1940-1975 period the 1$^\circ$ model is biased low and the 4$^\circ$ model is biased high. This leads to a better representation of the overall trend in the 1$^\circ$ model, as both models are biased low during 1996-2020.

We would also note that while we have trained the 4$^\circ$ model with the same hyperparameters as the 1$^\circ$ model for consistency, coarse model performance benefits from a larger embedding dimension, likely due to the increased subgrid activity at coarser resolution.

\subsection{Using CO$_2$ as an input feature}
\label{subsec:co2_sensitivity}

ACE2 uses both global-mean CO$_2$ concentration and spatially-varying SST as forcing when trained on either ERA5 or SHiELD. During the AMIP period, historical global-mean SSTs and CO$_2$ both increase with time, and the physical causality (i.e., gradual uptake of heat by the oceans due to increased radiative heating from elevated CO$_2$) may be difficult to learn from 6-hourly changes in the atmospheric and SST states. Here we evaluate the sensitivities of ACE2 to CO$_2$ specifically, by comparing ACE2 simulations with historical CO$_2$ to those where we set the concentration to a fixed value (1940 concentration of 307ppm), while retaining increasing SSTs.

\begin{figure}[t]
    \centering
    \includegraphics[width=\textwidth]{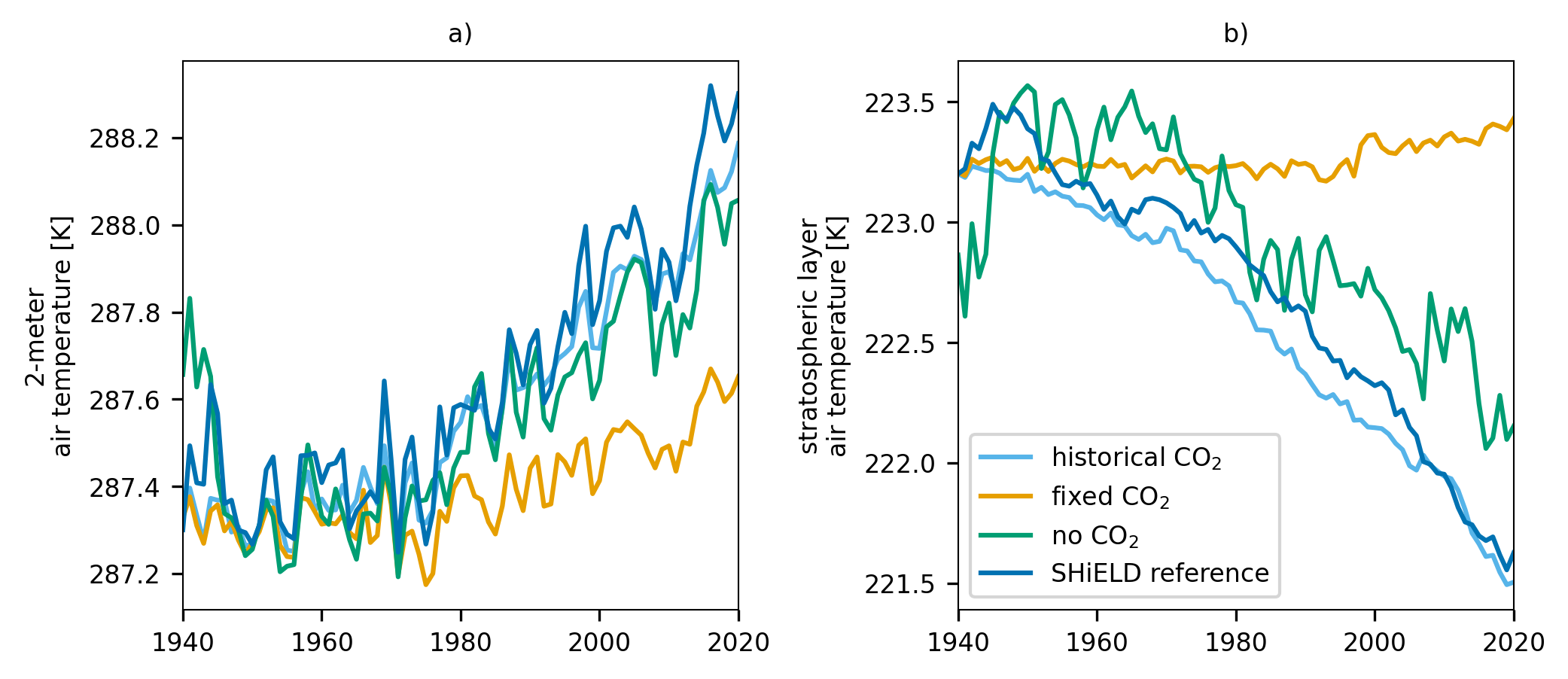}
    \caption{Global- and annual-mean (a) 2-meter air temperature and (b) level-0 (stratospheric) air temperature. Shown for the standard ACE2-SHiELD model ("historical CO$_2$"); the same model but with CO$_2$ concentrations fixed at the 1940 value ("fixed CO$_2$"); a version of ACE2-SHiELD trained without CO$_2$ as a feature ("no CO$_2$"); and the SHiELD reference data with historical SSTs and CO$_2$ forcing. The average of 3-member IC ensembles are shown.}
    \label{fig:co2_sensitivity}
\end{figure}

Figure \ref{fig:co2_sensitivity} shows that both near-surface and stratospheric global-mean temperature series in ACE2-SHiELD approximately match those in the reference dataset, when ACE2-SHiELD is forced with both historical SSTs and CO$_2$. There is both near-surface warming with polar amplification (Appendix \ref{appendix:temp_trend_maps}) and near-uniform stratospheric cooling. When holding CO$_2$ fixed, ACE2 no longer produces stratospheric cooling, as is expected physically \citep{Manabe1967}. However, it also loses much of the trend of near-surface warming, which is not expected. This is largely due to lack of warming over high-latitude land (not shown), despite evidence that such polar amplification should be driven largely by SST and sea ice coverage forcing \citep{Screen2012}.

In contrast, a version of ACE2 trained with SSTs but not CO$_2$ as forcing has global trends of near-surface warming and stratospheric cooling that somewhat underestimates these trends in the reference data, as well as excess inter-annual variability of stratospheric temperature. Thus using CO$_2$ as forcing with AMIP training datasets appears to improve the representation of some aspects of CO$_2$-induced trends, while introducing non-physical relationships in others. We lack SHiELD simulations forced by historical SSTs and fixed CO$_2$ (and vice versa), but these could be generated to augment ACE2's training data and test whether this improves the physical sensitivities of the emulator.

\section{Discussion}

This study demonstrates the feasibility of training a machine learning emulator to accurately generate atmospheric variability and forced responses from time scales of days to decades. ACE2 has a realistic global mean atmospheric response to increased sea surface temperature and CO$_2$. It generates realistic variability including the atmospheric response to El Ni\~no, the Madden Julian Oscillation, the geographic distribution of tropical cyclones and stratospheric polar vortex strength variability. By formulating ACE2 as an autoregressive model which simulates century-long trends through stepping forward 6 hours at a time, we can ensure physical consistency. Specifically, ACE2 exactly conserves dry air mass and moisture. Furthermore, by simulating climate as the average of explicitly resolved weather, intepretability is improved. As an example, the mechanisms by which ACE2 simulates the correct atmospheric response to El Ni\~no could be explored in a manner analogous to traditional numerical models.

Limitations of this work include the particular datasets used. For example, due to training on data corresponding to the last 80 years, we do not expect ACE2 to be able to properly simulate the response to strong climate change (e.g. a doubling of CO$_2$).  Furthermore, the SHiELD and ERA5 datasets both have shortcomings in accurately representing the true past conditions of the atmosphere. SHiELD is a coarse atmospheric model, and has biases in its global circulation. While the ERA5 dataset involves a data assimilation scheme to constrain its state to remain close to observations, fields such as the surface precipitation rate and radiative fluxes are not constrained and exhibit non-trivial biases with respect to satellite and station observations \citep{Urraca2018,Hersbach2020}. Furthermore, ACE2 itself does not accurately represent the expected atmospheric response to increasing sea surface temperature while keeping CO$_2$ fixed (Section~\ref{subsec:co2_sensitivity}) suggesting a need to encode the impacts of greenhouse gases in a more sophisticated manner. In addition, ACE2 does not exactly conserve global atmospheric energy, because it has a more complex budget equation and has significant non-conservation errors in atmospheric models such as SHiELD.

Future work will train ACE2 on SHiELD simulations spanning a wider range of CO$_2$ concentrations. In addition, the ability to simulate additional components of the climate system, such as ocean and sea ice, is a basic requirement for a useful climate model emulator.

\section{Methods}

\subsection{Versioning nomenclature}
We use the following nomenclature to distinguish between versions of the ACE model. ACE-climSST refers to the first version of ACE \citep{WattMeyer2023} which was trained on a dataset produced by forcing an atmospheric model with annually-repeating climatological SSTs and otherwise fixed external forcing. In this study, we introduce ACE2, which has an increased parameter count and updated loss function, introduces hard physical constraints on mass and moisture and uses a new checkpoint selection strategy in training, among other changes described below. We present results from training ACE2 on two distinct datasets, described in the next section. To distinguish these models, we will describe them as ACE2-SHiELD and ACE2-ERA5 respectively.

\subsection{Datasets}
Two datasets are used as targets for emulation (Table~\ref{table:datasets}). The first is output from the SHiELD atmospheric model \citep{Harris2020} at C96 (approximately 100$\,$km) resolution forced by observed sea surface temperatures and greenhouse gases from the 1940-2021 period. The latter is the ERA5 reanalysis dataset \citep{Hersbach2020} from 1940-2022. Other than their sources, the datasets are the same in terms of variable set (see Table~\ref{table:variables}) and resolution. ACE2, like ACE, combines the model-level fields for air temperature, specific total water and horizontal winds into eight vertical layers. The 2D prognostic variables are surface pressure, surface temperature over land and sea-ice, 2-meter air temperature and specific humidity and 10-meter horizontal winds. These latter near-surface variables are new additions compared to ACE \citep{WattMeyer2023} and are included due to their human impact relevance and importance for ocean coupling. Additional variables, used as diagnostics (outputs) only are the top-of-atmosphere and surface radiative fluxes, surface latent and sensible heat fluxes, surface precipitation rate and, for convenience, the 500hPa geopotential height and 850hPa air temperature. Finally, forcing variables (i.e. inputs only) are sea surface temperature, global-mean carbon dioxide (broadcast to a spatially uniform global field), incoming solar radiation at the top of atmosphere, land fraction, ocean fraction, sea ice fraction and surface topography. The use of carbon dioxide as a forcing input is a change from \cite{WattMeyer2023}.

The reference data is horizontally interpolated to the 
 1° Gaussian grid and the 6 hour temporal resolution used by ACE2 and ACE. All flux variables (e.g. radiative fluxes, precipitation) are time-averaged over the 6-hour intervals in order to enable exact evaluation of atmospheric budgets at the 6-hourly time resolution.  

\begin{table}
  \caption{Datasets used in this study. ERA5 is a reanalysis product, here coarsened to 1° horizontal resolution \citep{Hersbach2020}. SHiELD is an approximately 100 km resolution global atmospheric model which was forced by historical sea surface temperatures \citep{Harris2020}. For the SHiELD dataset, data is available from two ensemble members initialized from slightly different initial conditions on October 1, 1939, doubling the number of samples available.}
  \label{table:datasets}
  \centering
  \begin{tabular}{llll}
    \toprule
    Name   & Train period    & Validation period & Test period        \\
    \midrule
    ERA5    & 1940-1995, 2011-2019, 2021-2022 & 1996-2000 & 2001-2010, 2020 \\
    SHiELD  & 1940-1995, 2011-2021 & 1996-2000 & 2001-2010 \\
    \bottomrule
  \end{tabular}
\end{table}

\paragraph{SHiELD}

To generate multiple physics-based realizations of climate forced by historically observed sea surface temperatures, sea ice, and carbon dioxide, we make use of the public version of the SHiELD model developed at the Geophysical Fluid Dynamics Laboratory (GFDL) \citep{Harris2020}.  This is GFDL's developmental version of the FV3GFS model used in \cite{WattMeyer2023}.  The two models share a significant fraction of their code, the most notable difference being that SHiELD computes all microphysical updates every vertical remapping timestep within the dynamical core, rather than splitting the microphysical updates between the dynamical core and the physics \citep{Harris2020, Zho2022}.  

We run SHiELD at two horizontal resolutions, C96 (roughly 100~km) and C24 (roughly 400~km), with 79 vertical levels; C96 simulation output forms the basis of our target dataset, while C24 simulation output serves as a computationally inexpensive physics-based baseline.  Other than those related to horizontal resolution and convection—here we use the latest versions of both the shallow and deep convection schemes—we configure the parameters of the model following how they were configured in the C3072 (roughly 3~km) resolution X-SHiELD runs of \cite{Che2022}.  Note that no special tuning was attempted to help the climate of SHiELD better match observations when run at coarser resolution.  However we reduced a parameter controlling the strength of the mountain blocking scheme in the C24 configuration to help its climate, particularly the near-surface temperature over land, better match that of the C96 configuration based on the scheme's empirical sensitivity to resolution (J. Alpert and F. Yang, personal communication, August 9, 2019).

At each horizontal resolution, we run two identically forced simulations over 1940-2021, but with different initial conditions.  The initial conditions are generated by running a spin up simulation starting from GFS analysis for 2020-01-01 with 1930-01-01 forcing data for 117 months to 1939-10-01, outputting daily restart files from the last month.  This roughly 10-year period is meant to allow the model to adjust to the historical forcing after being initialized with present-day atmospheric conditions; the timescale is mainly limited by the time it takes stratospheric water vapor to equilibrate.  The restart files from 1939-09-30 and 1939-10-01 represent the state with which we start the two ensemble members on 1939-10-01, providing three months of spin up time prior to 1940-01-01 to allow the model states to meteorologically diverge.  A similar approach was used to generate initial conditions in the coupled model ensemble context in \cite{Des2012}.  We run the simulations until 2021-12-16T12:00:00, the last available time in our reference SST and sea ice dataset.  

The historical SST and sea ice concentration data come from that used to force historical AMIP CMIP6 simulations \citep{Tay2000,Eyring2016,Dur2022} and are provided on a 1° regular latitude-longitude grid as a monthly time series; space and time interpolation occurs online at the time of prescription within SHiELD.  We prescribe carbon dioxide as a time series of annual and global means, with data prior to 2015 coming from that used for CMIP6 \citep{Mei2017} and data after coming from the NOAA Global Monitoring Laboratory \citep{Con1994}; in these runs we assume CO$_2$ is well-mixed (i.e. globally uniform).

Data from these simulations is output on the model native cubed-sphere grid at 6-hourly intervals.  We make use of GFDL's \texttt{fregrid} tool \citep{fregrid2024} to conservatively regrid the model state to a Gaussian grid.  In the case of C96 data this is a 1° grid, and in the case of C24 data this is a 4° grid. Similar to a regular latitude-longitude grid, a Gaussian grid provides increased resolution in the polar regions, which means that with a conservative regridding approach the original cubed-sphere grid cell edges in these regions are resolved with high fidelity. As in \cite{WattMeyer2023}, we perform a spherical harmonic transform (SHT) round trip on all but the surface type fraction variables in the regridded output to smooth these sharp boundaries, which otherwise produce artifacts under spherical harmonic transforms.  Finally we coarsen vertically resolved fields from the native 79 vertical layers to ACE's 8 layers with mass-weighted averages.

\paragraph{ERA5}
We use the ERA5 reanalysis dataset spanning 1940-2022. Our version of the dataset---at 1° horizontal resolution and with 8 terrain-following vertical layers---is derived from the native dataset on 137 model layers and stored in terms of spherical harmonic coefficients or on a reduced Gaussian grid, depending on the variable. It was computed from the version of ERA5 hosted by Google Research (https://github.com/google-research/arco-era5; \cite{Carver2023}). Routines from the MetView package \citep{metview} were used for the regridding. To the extent possible, data was regridded and vertically coarsened to match the SHiELD dataset's horizontal and vertical coordinate. Unlike the SHiELD dataset, no spherical harmonic round trip was performed on the data.

\subsection{Training}
\label{subsec:training}
\paragraph{Architecture}

The Spherical Fourier Neural Operator (SFNO) architecture is used \citep{Bonev2023}. This is a neural operator type architecture well suited to data on the sphere. This is the same architecture used in \cite{WattMeyer2023}. The only difference in configuration of the SFNO from version 1 of ACE is that the embedding dimension is increased from 256 to 384 for ACE2. In addition, a corrector imposing physical constraints is included as part of the model architecture, as described in the next section.

\paragraph{Hard physical constraints}
\label{sec:constraints-methods}

In our previous work, we found global mean surface pressure drifted unrealistically \citep[c.f. Figure 9 of][]{WattMeyer2023}. And while the model very nearly obeyed the column-wise conservation of moisture without an explicit penalty or constraint, there were still small violations of this budget and the global mean moisture budget was violated by up to 0.1 mm/day at individual time steps \citep[c.f. Figure 11 of][]{WattMeyer2023}. Here we describe how we enforce hard physical constraints to eliminate these budget violations. The following equations define the budgets which we desire to impose. First, conservation of global dry air mass:

\begin{equation}
    \label{eq:psdry}
    \langle {p_s^{dry}(t + \Delta t) } \rangle = \langle {p_s^{dry}(t) } \rangle
\end{equation}

where $p_s^{dry}(t) = p_s(t) - g TWP(t)$ is the surface pressure due to dry air, $TWP(t) = \frac{1}{g}\int_0^{p_s}q(t,p) dp$ is the total water path, $\Delta t$ is the forward time step of the machine learning model and angled brackets $\langle \rangle$ represent the area-weighted global average. Next, the conservation of column-integrated moisture:

\begin{equation}
\label{eq:columnmoisture}
    \frac{TWP(t + \Delta t) - TWP(t)}{\Delta t} = E(t) - P(t) + \left. \frac{\partial TWP}{\partial t}\right |_{adv}(t)
\end{equation}

where $E(t)$ is the evaporation rate, computed as $LHF(t)/L_v$, $P$ is the precipitation rate and $\left. \frac{\partial TWP}{\partial t}\right |_{adv}$ is the tendency of total water path due to advection, which is directly predicted by the machine learning model (see also Table~\ref{table:variables}). Note that all of the terms on the right hand side of Equation~\ref{eq:columnmoisture} represent time averages between $t$ and $t + \Delta t$. Finally, we have the constraints on global moisture:

\begin{equation}
  \left  \langle \left. \frac{\partial TWP}{\partial t}\right |_{adv}(t) \right \rangle = 0
\end{equation}
and by implication

\begin{equation}
\label{eq:globalmoisture}
    \left \langle \frac{TWP(t + \Delta t) - TWP(t)}{\Delta t} \right \rangle = \left \langle E(t) - P(t) \right \rangle.
\end{equation}

We enforce these physical constraints on the model by including a physical corrector module within the optimized model. This module applies the following corrections to ensure the constraints are satisfied:

\begin{enumerate}
    \item Moisture, precipitation rate, and radiative fluxes are all made to be positive by setting any negative values to zero.
    \item A globally-constant surface pressure adjustment ensures total dry air mass is conserved:

        \begin{equation}
           \label{eq:pscorrection}
           p_s'(t) = p_s(t) - \langle {p_s^{dry}(t) - p_s^{dry}(t-1)} \rangle
        \end{equation}
        
    \item Precipitation rate is multiplied by a constant to conserve global mean moisture:
        
        \begin{equation}
           \label{eq:twp}
           P'(t) = \frac{P(t)}{\langle P(t) \rangle} \langle E(t) - \frac{TWP(t) - TWP(t-1)}{\Delta t}  \rangle,
        \end{equation}
        
        where $P'(t)$ is the corrected precipitation rate at time $t$, $P(t)$ is the precipitation prior to this correction, and $E = LHF(t) / L_v$ is the evaporation rate.
    
    \item Exact conservation of column moisture is attained by deriving advective flux as residual from the adjusted TWP tendency, E and P:

        \begin{equation}
           \label{eq:advective}
           \left. \frac{\partial TWP}{\partial t}\right |_{adv}' = \frac{TWP(t) - TWP(t-1)}{\Delta t} - (E(t) - P'(t)),
        \end{equation}
        
        where $\left. \frac{\partial TWP}{\partial t}\right |_{adv}'$ represents the corrected tendency of total water path due to advection. 
    
\end{enumerate}

We avoid introducing model bias through these corrections by applying them before computing the loss. For this reason, these constraints can be considered to be part of the model architecture. The order of these adjustments is such that later corrections will not invalidate earlier corrections. These corrections are applied, by necessity, to the data in physical units instead of in normalized units.

\paragraph{Data Normalization}
For the inputs and outputs of the SFNO module, data is normalized using standard scaling. Means and standard deviations are computed over latitude, longitude and time without any area weighting.  For normalization before the loss function is computed, prognostic variables are scaled to harmonize their typical difference between time steps, i.e. we use ``residual'' scaling (see Appendix H of \cite{WattMeyer2023}). Specifically, for a field $a(t, \phi, \lambda)$ which depends on time, latitude and longitude, the standard deviation of $a(t+\Delta t, \phi, \lambda) - a(t, \phi, \lambda)$ over time and space is used for normalization. Diagnostic variables are normalized for the loss function using standard scaling.

For the ERA5 dataset, normalization statistics were computed over the period 1990-2020 for which this reanalysis is most reliable. For the SHiELD dataset, they were computed over 1940-2021.

\paragraph{Loss Function}
The loss function is the mean squared error over all outputs. Prognostic outputs are normalized using residual scaling as described in previous section while diagnostic outputs are normalized using standard full field scaling. The loss is summed over two autoregressive forward 6-hour steps. In addition, some variables are given an additional weighting (Table~\ref{table:lossweights}). Variables which were downweighted are ones which showed signs of overfitting (that is, increasing 6-hour RMSE on validation data late in training) without the downweighting. Variables which are upweighted are diagnostic variables, which would otherwise contribute relatively little (<0.5\%) to the loss function that is averaged across 50 outputs.

\begin{table}
  \caption{Custom weights applied to variables when computing loss function. Output variables which are not listed here are given a weight of 1. Variables are defined in Table~\ref{table:variables}.}
  \label{table:lossweights}
  \centering
  \begin{tabular}{lc}
    \toprule
    Name   & Weight        \\
    \midrule
    $T_0$, $T_1$, $u_0$, $v_0$, $q_0$, $q_2$, $q_{2m}$, $P$, $\left. \frac{\partial TWP}{\partial t}\right |_{adv}$   & 0.5 \\
    $q_1$  & 0.25 \\
    DLWRF$_{sfc}$, USWRF$_{sfc}$, DSWRF$_{sfc}$, USWRF$_{toa}$ & 2 \\
    ULWRF$_{sfc}$, $T_{850}$ & 5 \\
    $Z_{500}$ & 10 \\
    \bottomrule
  \end{tabular}
\end{table}

\paragraph{Checkpoint selection based on climate skill}
Since the loss function used here is based on 12-hour forecast skill over two 6-hourly autoregressive steps, it is not guaranteed that a lower loss will lead to small long-term (e.g. 10-year averaged) climate biases. Since our priority in this work is accurate representation of climate statistics, we therefore define a selection criteria to choose a checkpoint with the smallest time-averaged biases. The criteria is the channel-mean global RMSE of time-means. Specifically:
\begin{equation}
\label{eq:channel_and_time_mean_RMSE}
    \alpha = \frac{1}{C}\sum_{c=1}^C \sqrt{\sum_{\phi, \lambda} w_{\phi,\lambda} \left ( \overline{y_c(t, \phi, \lambda) - \hat{y}_c(t, \phi, \lambda )} \right )^2}
\end{equation}
where $c$ is an index for output channel (i.e. the prognostic and diagnostic variables), $w_\phi$ is an averaging weight proportional to area of grid cell centered at $\phi, \lambda$. The $y_c(t, \phi, \lambda)$ is the normalized true value at a particular time and location, while $\hat{y}_c$ is the normalized model prediction for the corresponding time, from a simulation initialized at some previous time. The overbar $\overline{\cdot}$ is a time- and ensemble-average.

In practice, $\alpha$ is computed once per epoch during training from an ensemble of eight 5-year long simulations, initialized at evenly spaced intervals across 1996, the start of the validation period (Table~\ref{table:datasets}). In addition to choosing a best checkpoint from within a training run, we perform an ensemble of four training runs which differ only in the initialization of model parameters. For each training run, we choose a checkpoint based on minimizing $\alpha$ across epochs. After these training runs were completed, we found that doing inference runs over a wider span of forcing data led to a better estimate of the climate skill of a given model. Therefore to choose a checkpoint across the four training runs, we performed twelve 5-year inference runs, initialized once every 5 years starting on 1 January 1940, spanning the training and validation periods, but not overlapping with the held out test period. Additionally, for this comparison we downweighted the contribution of $q_0$ to the calculation of $\alpha$ by a factor of 10, since our poor skill in predicting the time-mean of this variable otherwise dominated $\alpha$. Then the checkpoint across the four random seeds was chosen according to this new criteria. Appendix~\ref{appendix:discussion_seed_variability} shows the variability of $\alpha$ through training and across the four random seeds.

\subsection{Evaluation metrics}

To evaluate time-mean climate skill, we compute the global RMSE of the time-mean for an individual variable $y$ as:
\begin{equation}
\label{eq:time_mean_RMSE}
     \sqrt{\sum_{\phi, \lambda} w_{\phi,\lambda} \left ( \overline{y(t, \phi, \lambda) - \hat{y}(t, \phi, \lambda )} \right )^2}
\end{equation}
where $w_{\phi,\lambda}$ is an area weight that sums to 1 and $\hat{y}$ is the predicted value and the overline represents a time average. For global- and annual-mean series of a given output variable, we also compute an R$^2$ of the predicted series against a reference series of that variable: 

\begin{equation}
    R^2 = 1 - \frac{SS_{error}}{SS_{reference}}
\end{equation}
where:
\begin{equation}
\label{eq:SS_error}
    SS_{error} = \sum^{n_{years}}_{i_{year}=1}{(\hat{y}_{i_{year}} - y_{i_{year}})}^2
\end{equation}
and 
\begin{equation}
\label{eq:SS_reference}
    SS_{reference} = \sum^{n_{years}}_{i_{year}=1}{(y_{i_{year}} - \bar{y})}^2.
\end{equation}

Here $\hat{y}$ and $y$ are predicted and reference variable values, and $\bar{y}$ is the average over the time period. Thus $R^2$ reflects the model's combined ability to capture long-term means and trends as well as shorter-term inter-annual variability. 

To characterize the atmospheric response to El Ni\~no-Southern Oscillation (ENSO) we compute a regression coefficient of variables against the historical Ni\~no 3.4 index \citep{Trenberth1997} as computed from the CMIP6 AMIP SST dataset \citep{Tay2000, Eyring2016, Dur2022}. The coefficient is $\beta_1$ of a linear regression:

\begin{equation}
\label{eq:nino_coefficient}
    \hat{y} = \beta_1 I_{Ni\Tilde{n}o34} + \beta_0    
\end{equation}

where $I_{Ni\Tilde{n}o34}$ is the 3-monthly centered running mean of SSTs in the Ni\~no 3.4 region, after being nearest-neighbor interpolated to the 6-hourly time frequency of data. This produces a map of the response of a particular variable to seasonally-varying ENSO states. We compare the predicted response against a reference dataset response by computing the global area-weighted RMS difference between the response maps. This also allows for computing the variability of the SHiELD reference dataset's atmospheric response to ENSO, as the difference between the response maps of its two initial conditions.

\subsection{Computational cost}
Training duration for each model is approximately 4.5 days on eight NVIDIA H100-80GB-HBM3 GPUs. For each dataset, four models were trained with the same hyperparameters and differing only in parameter initialization (see ``Checkpoint selection based on climate skill'' section above) quadrupling the overall cost. The cost of doing inference with ACE2 and the reference SHiELD model is shown in Table~\ref{table:inferencecost}. Comparing ACE2 and C96 SHiELD, which have the same horizontal resolution, ACE2 is about 100 times faster and 700 times less energy intensive. Even compared to C24 SHiELD, which has four times lower horizontal resolution, ACE2 uses about 25 times less energy and is about 50 times faster.

\begin{table}
  \caption{Speed and energy cost of inference with ACE2 and the physics-based SHiELD model. ACE2 and C96 SHiELD both have about 1° horizontal resolution while C24 SHiELD has about 4° resolution but is still an order of magnitude more energy-intensive than ACE2.}
  \label{table:inferencecost}
  \centering
  \begin{tabular}{p{1.9cm}p{1.9cm}p{2.9cm}p{5.5cm}}
    \toprule
    Model   & Simulated years per day & Energy cost per simulated year [Wh] & Hardware \\
    \midrule
    ACE2        & 1500 & 11.2 & 1 NVIDIA H100-80GB-HBM \\
    C24 SHiELD  & 22.1 & 300 & 54 cores on 1 AMD EPYC 7H12 \\
    C96 SHiELD  & 11.4 & 8250 & 864 cores on 14 AMD EPYC 7H12 \\
    \bottomrule
  \end{tabular}
\end{table}


%

\section{Data availability}
The ERA5 dataset \citep{Hersbach2020} is available from the Copernicus Climate Data Store (https://cds.climate.copernicus.eu/). The processed version of the dataset used to train ACE2-ERA5 is available on a public requester-pays Google Cloud Storage bucket at \texttt{gs://ai2cm-public-requester-pays/2024-11-13-ai2-climate-emulator-v2-amip/data/} \texttt{era5-1deg-1940-2022.zarr} (about 1.5TiB). Similarly, the SHiELD dataset used to train ACE2-SHiELD is available at \texttt{gs://ai2cm-public-requester-pays/2024-11-13-ai2-climate-emulator-v2-amip/data/} \texttt{c96-1deg-shield} (about 3 TiB).

\section{Code availability}
The code used for data processing, model training, inference and evaluation is available at \href{https://github.com/ai2cm/ace}{https://github.com/ai2cm/ace}. The trained ACE2-ERA5 model checkpoint is available at \href{https://huggingface.co/allenai/ACE2-ERA5}{https://huggingface.co/allenai/ACE2-ERA5}.

\begin{ack}
We acknowledge NOAA's Geophysical Fluid Dynamics Laboratory for providing the computing resources used to perform the reference SHiELD simulations. This research used resources of NERSC, a U.S. Department of Energy Office of Science User Facility located at Lawrence Berkeley National Laboratory, using NERSC award BER-ERCAP0026743. We acknowledge ECMWF for generating and providing the ERA5 dataset.
\end{ack}

\bibliography{main}

\begin{thebibliography}{67}
\providecommand{\natexlab}[1]{#1}
\expandafter\ifx\csname urlstyle\endcsname\relax
  \providecommand{\doi}[1]{doi:\discretionary{}{}{}#1}\else
  \providecommand{\doi}{doi:\discretionary{}{}{}\begingroup \urlstyle{rm}\Url}\fi

\bibitem[{Ahn et~al.(2020)Ahn, Kim, Kang, Lee, Sperber et~al.}]{Ahn2020}
Min‐Seop Ahn, Daehyun Kim, Daehyun Kang, Jiwoo Lee, Kenneth~R. Sperber, et~al.
\newblock MJO Propagation Across the Maritime Continent: Are CMIP6 Models Better Than CMIP5 Models?
\newblock \emph{Geophysical Research Letters}, 47(11), 2020.
\newblock \doi{10.1029/2020gl087250}.

\bibitem[{Anstey et~al.(2022)Anstey, Osprey, Alexander, Baldwin, Butchart et~al.}]{Anstey2022}
James~A. Anstey, Scott~M. Osprey, Joan Alexander, Mark~P. Baldwin, Neal Butchart, et~al.
\newblock Impacts, processes and projections of the quasi-biennial oscillation.
\newblock \emph{Nature Reviews Earth Environment}, 3(9):588–603, 2022.
\newblock \doi{10.1038/s43017-022-00323-7}.

\bibitem[{Baldwin and Dunkerton(2001)}]{Baldwin2001}
Mark~P. Baldwin and Timothy~J. Dunkerton.
\newblock Stratospheric Harbingers of Anomalous Weather Regimes.
\newblock \emph{Science}, 294(5542):581–584, 2001.
\newblock \doi{10.1126/science.1063315}.

\bibitem[{Beucler et~al.(2024)Beucler, Gentine, Yuval, Gupta, Peng et~al.}]{Beucler2024}
Tom Beucler, Pierre Gentine, Janni Yuval, Ankitesh Gupta, Liran Peng, et~al.
\newblock Climate-invariant machine learning.
\newblock \emph{Science Advances}, 10(6):eadj7250, 2024.
\newblock \doi{10.1126/sciadv.adj7250}.

\bibitem[{Bhatia et~al.(2019)Bhatia, Vecchi, Knutson, Murakami, Kossin et~al.}]{Bhatia2019}
Kieran~T. Bhatia, Gabriel~A. Vecchi, Thomas~R. Knutson, Hiroyuki Murakami, James Kossin, et~al.
\newblock Recent increases in tropical cyclone intensification rates.
\newblock \emph{Nature Communications}, 10(1), 2019.
\newblock \doi{10.1038/s41467-019-08471-z}.

\bibitem[{Bi et~al.(2023{\natexlab{a}})Bi, Xie, Zhang, Chen, Gu et~al.}]{PanguWeather}
Kaifeng Bi, Lingxi Xie, Hengheng Zhang, Xin Chen, Xiaotao Gu, et~al.
\newblock Accurate medium-range global weather forecasting with 3D neural networks.
\newblock \emph{Nature}, 619(7970):533--538, 2023{\natexlab{a}}.
\newblock \doi{10.1038/s41586-023-06185-3}.

\bibitem[{Bi et~al.(2023{\natexlab{b}})Bi, Xie, Zhang, Chen, Gu et~al.}]{Bi2023}
Kaifeng Bi, Lingxi Xie, Hengheng Zhang, Xin Chen, Xiaotao Gu, et~al.
\newblock Accurate medium-range global weather forecasting with 3D neural networks.
\newblock \emph{Nature}, 619(7970):533–538, 2023{\natexlab{b}}.
\newblock \doi{10.1038/s41586-023-06185-3}.

\bibitem[{Bonev et~al.(2023)Bonev, Kurth, Hundt, Pathak, Baust et~al.}]{Bonev2023}
Boris Bonev, Thorsten Kurth, Christian Hundt, Jaideep Pathak, Maximilian Baust, et~al.
\newblock Spherical Fourier Neural Operators: Learning Stable Dynamics on the Sphere.
\newblock \emph{Proceedings of the 40th International Conference on Machine Learning (ICML)}, 2023.
\newblock \doi{10.48550/ARXIV.2306.03838}.

\bibitem[{Brajard et~al.(2020)Brajard, Carrassi, Bocquet, and Bertino}]{Brajard2020}
Julien Brajard, Alberto Carrassi, Marc Bocquet, and Laurent Bertino.
\newblock Combining data assimilation and machine learning to emulate a dynamical model from sparse and noisy observations: A case study with the Lorenz 96 model.
\newblock \emph{Journal of Computational Science}, 44:101171, 2020.
\newblock \doi{10.1016/j.jocs.2020.101171}.

\bibitem[{Carver and Merose(2023)}]{Carver2023}
Robert~W. Carver and Alex Merose.
\newblock {ARCO-ERA5: An Analysis-Ready Cloud-Optimized Reanalysis Dataset}.
\newblock 22nd Conf. on AI for Env. Science, Denver, CO, Amer. Meteo. Soc., 2023.

\bibitem[{Chen et~al.(2024)Chen, Zhong, Li, Wu, Lu et~al.}]{Chen2024}
Lei Chen, Xiaohui Zhong, Hao Li, Jie Wu, Bo~Lu, et~al.
\newblock A machine learning model that outperforms conventional global subseasonal forecast models.
\newblock \emph{Nature Communications}, 15(1), 2024.
\newblock \doi{10.1038/s41467-024-50714-1}.

\bibitem[{Chen et~al.(2023)Chen, Zhong, Zhang, Cheng, Xu et~al.}]{Chen2023}
Lei Chen, Xiaohui Zhong, Feng Zhang, Yuan Cheng, Yinghui Xu, et~al.
\newblock FuXi: a cascade machine learning forecasting system for 15-day global weather forecast.
\newblock \emph{npj Climate and Atmospheric Science}, 6(1), 2023.
\newblock \doi{10.1038/s41612-023-00512-1}.

\bibitem[{Cheng et~al.(2022)Cheng, Harris, Bretherton, Merlis, Bolot et~al.}]{Che2022}
Kai-Yuan Cheng, Lucas Harris, Christopher Bretherton, Timothy~M. Merlis, Maximilien Bolot, et~al.
\newblock Impact of {{Warmer Sea Surface Temperature}} on the {{Global Pattern}} of {{Intense Convection}}: {{Insights From}} a {{Global Storm Resolving Model}}.
\newblock \emph{Geophysical Research Letters}, 49(16):e2022GL099796, 2022.
\newblock \doi{10.1029/2022GL099796}.

\bibitem[{Clark et~al.(2022)Clark, Brenowitz, Henn, Kwa, McGibbon et~al.}]{Clark2022}
Spencer~K. Clark, Noah~D. Brenowitz, Brian Henn, Anna Kwa, Jeremy McGibbon, et~al.
\newblock Correcting a 200 km Resolution Climate Model in Multiple Climates by Machine Learning From 25 km Resolution Simulations.
\newblock \emph{Journal of Advances in Modeling Earth Systems}, 14(9), 2022.
\newblock \doi{10.1029/2022ms003219}.

\bibitem[{Claussen et~al.(2002)Claussen, Mysak, Weaver, M., Fichefet et~al.}]{Claussen2002}
M.~Claussen, L.~Mysak, A.~Weaver, Crucifix M., T.~Fichefet, et~al.
\newblock Earth system models of intermediate complexity: closing the gap in the spectrum of climate system models.
\newblock \emph{Climate Dynamics}, 18(7):579–586, 2002.
\newblock \doi{10.1007/s00382-001-0200-1}.

\bibitem[{Collins et~al.(2004)Collins, Rasch, Boville, McCaa, Williamson et~al.}]{Collins2004}
William Collins, Philip Rasch, Byron Boville, James McCaa, David Williamson, et~al.
\newblock Description of the {NCAR Community Atmosphere Model (CAM 3.0)}.
\newblock Technical report, {UCAR/NCAR}, 2004.
\newblock \doi{10.5065/D63N21CH}.

\bibitem[{Conway et~al.(1994)Conway, Tans, Waterman, Thoning, Kitzis et~al.}]{Con1994}
Thomas~J. Conway, Pieter~P. Tans, Lee~S. Waterman, Kirk~W. Thoning, Duane~R. Kitzis, et~al.
\newblock Evidence for Interannual Variability of the Carbon Cycle from the {{National Oceanic}} and {{Atmospheric Administration}}/{{Climate Monitoring}} and {{Diagnostics Laboratory Global Air Sampling Network}}.
\newblock \emph{Journal of Geophysical Research: Atmospheres}, 99(D11):22831--22855, 1994.
\newblock \doi{10.1029/94JD01951}.

\bibitem[{Cresswell-Clay et~al.(2024)Cresswell-Clay, Liu, Durran, Liu, Espinosa et~al.}]{Cresswell2024}
Nathaniel Cresswell-Clay, Bowen Liu, Dale Durran, Andy Liu, Zachary~I. Espinosa, et~al.
\newblock A Deep Learning Earth System Model for Stable and Efficient Simulation of the Current Climate.
\newblock 2024.
\newblock \doi{10.48550/ARXIV.2409.16247}.

\bibitem[{Deser et~al.(2012)Deser, Phillips, Bourdette, and Teng}]{Des2012}
Clara Deser, Adam Phillips, Vincent Bourdette, and Haiyan Teng.
\newblock Uncertainty in Climate Change Projections: The Role of Internal Variability.
\newblock \emph{Climate Dynamics}, 38(3):527--546, 2012.
\newblock \doi{10.1007/s00382-010-0977-x}.

\bibitem[{Duncan et~al.(2024)Duncan, Wu, Golaz, Caldwell, Watt‐Meyer et~al.}]{Duncan2024}
James P.~C. Duncan, Elynn Wu, Jean‐Christophe Golaz, Peter~M. Caldwell, Oliver Watt‐Meyer, et~al.
\newblock Application of the AI2 Climate Emulator to E3SMv2’s Global Atmosphere Model, With a Focus on Precipitation Fidelity.
\newblock \emph{Journal of Geophysical Research: Machine Learning and Computation}, 1(3), 2024.
\newblock \doi{10.1029/2024jh000136}.

\bibitem[{Durack et~al.(2022)Durack, Taylor, {Po-Chedley}, and Doutriaux}]{Dur2022}
Paul~J. Durack, Karl~E. Taylor, Stephen {Po-Chedley}, and Charles Doutriaux.
\newblock {{amipbcs}} - {{AMIP}} Dataset Prepared for {{input4MIPS}}.
\newblock 2022.

\bibitem[{Elsner et~al.(2008)Elsner, Kossin, and Jagger}]{Elsner2008}
James~B. Elsner, James~P. Kossin, and Thomas~H. Jagger.
\newblock The increasing intensity of the strongest tropical cyclones.
\newblock \emph{Nature}, 455(7209):92–95, 2008.
\newblock \doi{10.1038/nature07234}.

\bibitem[{Eyring et~al.(2016)Eyring, Bony, Meehl, Senior, Stevens et~al.}]{Eyring2016}
Veronika Eyring, Sandrine Bony, Gerald~A. Meehl, Catherine~A. Senior, Bjorn Stevens, et~al.
\newblock Overview of the Coupled Model Intercomparison Project Phase 6 (CMIP6) experimental design and organization.
\newblock \emph{Geoscientific Model Development}, 9(5):1937–1958, 2016.
\newblock \doi{10.5194/gmd-9-1937-2016}.

\bibitem[{Golaz et~al.(2022)Golaz, Van~Roekel, Zheng, Roberts, Wolfe et~al.}]{Golaz2022}
Jean‐Christophe Golaz, Luke~P. Van~Roekel, Xue Zheng, Andrew~F. Roberts, Jonathan~D. Wolfe, et~al.
\newblock The DOE E3SM Model Version 2: Overview of the Physical Model and Initial Model Evaluation.
\newblock \emph{Journal of Advances in Modeling Earth Systems}, 14(12), 2022.
\newblock \doi{10.1029/2022ms003156}.

\bibitem[{Guan et~al.(2024)Guan, Arcomano, Chattopadhyay, and Maulik}]{Guan2024}
Haiwen Guan, Troy Arcomano, Ashesh Chattopadhyay, and Romit Maulik.
\newblock LUCIE: A Lightweight Uncoupled ClImate Emulator with long-term stability and physical consistency for O(1000)-member ensembles.
\newblock 2024.
\newblock \doi{10.48550/ARXIV.2405.16297}.

\bibitem[{Hagos et~al.(2011)Hagos, Leung, and Dudhia}]{Hagos2011}
Samson Hagos, L.~Ruby Leung, and Jimy Dudhia.
\newblock Thermodynamics of the Madden–Julian Oscillation in a Regional Model with Constrained Moisture.
\newblock \emph{Journal of the Atmospheric Sciences}, 68(9):1974–1989, 2011.
\newblock \doi{10.1175/2011jas3592.1}.

\bibitem[{Harris et~al.(2020)Harris, Zhou, Lin, Chen, Chen et~al.}]{Harris2020}
Lucas Harris, Linjiong Zhou, Shian‐Jiann Lin, Jan‐Huey Chen, Xi~Chen, et~al.
\newblock GFDL SHiELD: A Unified System for Weather‐to‐Seasonal Prediction.
\newblock \emph{Journal of Advances in Modeling Earth Systems}, 12(10), 2020.
\newblock \doi{10.1029/2020ms002223}.

\bibitem[{Hatfield et~al.(2021)Hatfield, Chantry, Dueben, Lopez, Geer et~al.}]{Hatfield2021}
Sam Hatfield, Matthew Chantry, Peter Dueben, Philippe Lopez, Alan Geer, et~al.
\newblock Building Tangent‐Linear and Adjoint Models for Data Assimilation With Neural Networks.
\newblock \emph{Journal of Advances in Modeling Earth Systems}, 13(9), 2021.
\newblock \doi{10.1029/2021ms002521}.

\bibitem[{Hersbach et~al.(2020)Hersbach, Bell, Berrisford, Hirahara, Horányi et~al.}]{Hersbach2020}
Hans Hersbach, Bill Bell, Paul Berrisford, Shoji Hirahara, András Horányi, et~al.
\newblock The ERA5 global reanalysis.
\newblock \emph{Quarterly Journal of the Royal Meteorological Society}, 146(730):1999–2049, 2020.
\newblock \doi{10.1002/qj.3803}.

\bibitem[{Hodges et~al.(2017)Hodges, Cobb, and Vidale}]{Hodges2017}
Kevin Hodges, Alison Cobb, and Pier~Luigi Vidale.
\newblock How Well Are Tropical Cyclones Represented in Reanalysis Datasets?
\newblock \emph{Journal of Climate}, 30(14):5243–5264, 2017.
\newblock \doi{10.1175/jcli-d-16-0557.1}.

\bibitem[{Karlbauer et~al.(2024)Karlbauer, Cresswell‐Clay, Durran, Moreno, Kurth et~al.}]{Karlbauer2024}
Matthias Karlbauer, Nathaniel Cresswell‐Clay, Dale~R. Durran, Raul~A. Moreno, Thorsten Kurth, et~al.
\newblock Advancing Parsimonious Deep Learning Weather Prediction Using the HEALPix Mesh.
\newblock \emph{Journal of Advances in Modeling Earth Systems}, 16(8), 2024.
\newblock \doi{10.1029/2023ms004021}.

\bibitem[{Kay et~al.(2015)Kay, Deser, Phillips, Mai, Hannay et~al.}]{Kay2015}
J.~E. Kay, C.~Deser, A.~Phillips, A.~Mai, C.~Hannay, et~al.
\newblock The Community Earth System Model (CESM) Large Ensemble Project: A Community Resource for Studying Climate Change in the Presence of Internal Climate Variability.
\newblock \emph{Bulletin of the American Meteorological Society}, 96(8):1333–1349, 2015.
\newblock \doi{10.1175/bams-d-13-00255.1}.

\bibitem[{Kenneth et~al.(2019)Kenneth, Howard, James, Michael, and Carl}]{IBTRACS}
R.~Kenneth, J.~Howard, P.~James, C.~Michael, and J.~Carl.
\newblock International Best Track Archive for Climate Stewardship (IBTrACS) Project, Version 4.
\newblock 2019.
\newblock \doi{10.25921/82TY-9E16}.

\bibitem[{Kim et~al.(2009)Kim, Sperber, Stern, Waliser, Kang et~al.}]{Kim2009}
D.~Kim, K.~Sperber, W.~Stern, D.~Waliser, I.-S. Kang, et~al.
\newblock Application of MJO Simulation Diagnostics to Climate Models.
\newblock \emph{Journal of Climate}, 22(23):6413–6436, 2009.
\newblock \doi{10.1175/2009jcli3063.1}.

\bibitem[{Knapp et~al.(2010)Knapp, Kruk, Levinson, Diamond, and Neumann}]{Knapp2010}
Kenneth~R. Knapp, Michael~C. Kruk, David~H. Levinson, Howard~J. Diamond, and Charles~J. Neumann.
\newblock The International Best Track Archive for Climate Stewardship (IBTrACS): Unifying Tropical Cyclone Data.
\newblock \emph{Bulletin of the American Meteorological Society}, 91(3):363–376, 2010.
\newblock \doi{10.1175/2009bams2755.1}.

\bibitem[{Kochkov et~al.(2021)Kochkov, Smith, Alieva, Wang, Brenner et~al.}]{Kochov2021}
Dmitrii Kochkov, Jamie~A. Smith, Ayya Alieva, Qing Wang, Michael~P. Brenner, et~al.
\newblock Machine learning–accelerated computational fluid dynamics.
\newblock \emph{Proceedings of the National Academy of Sciences}, 118(21):e2101784118, 2021.
\newblock \doi{10.1073/pnas.2101784118}.

\bibitem[{Kochkov et~al.(2024)Kochkov, Yuval, Langmore, Norgaard, Smith et~al.}]{Kochkov2024}
Dmitrii Kochkov, Janni Yuval, Ian Langmore, Peter Norgaard, Jamie Smith, et~al.
\newblock Neural general circulation models for weather and climate.
\newblock \emph{Nature}, 632(8027):1060–1066, 2024.
\newblock \doi{10.1038/s41586-024-07744-y}.

\bibitem[{Kucharski et~al.(2013)Kucharski, Molteni, King, Farneti, Kang et~al.}]{Kucharski2013}
Fred Kucharski, Franco Molteni, Martin~P. King, Riccardo Farneti, In-Sik Kang, et~al.
\newblock On the Need of Intermediate Complexity General Circulation Models: A {\textquotedblleft}{SPEEDY}{\textquotedblright} Example.
\newblock \emph{Bulletin of the American Meteorological Society}, 94(1):25--30, 2013.
\newblock \doi{10.1175/bams-d-11-00238.1}.

\bibitem[{Lam et~al.(2023)Lam, Sanchez-Gonzalez, Willson, Wirnsberger, Fortunato et~al.}]{Lam2023}
Remi Lam, Alvaro Sanchez-Gonzalez, Matthew Willson, Peter Wirnsberger, Meire Fortunato, et~al.
\newblock Learning skillful medium-range global weather forecasting.
\newblock \emph{Science}, 382(6677):1416–1421, 2023.
\newblock \doi{10.1126/science.adi2336}.

\bibitem[{Mahesh et~al.(2024)Mahesh, Collins, Bonev, Brenowitz, Cohen et~al.}]{Manesh2024}
Ankur Mahesh, William Collins, Boris Bonev, Noah Brenowitz, Yair Cohen, et~al.
\newblock Huge Ensembles Part I: Design of Ensemble Weather Forecasts using Spherical Fourier Neural Operators.
\newblock 2024.
\newblock \doi{10.48550/ARXIV.2408.03100}.

\bibitem[{Manabe and Wetherald(1967)}]{Manabe1967}
Syukuro Manabe and Richard~T. Wetherald.
\newblock Thermal Equilibrium of the Atmosphere with a Given Distribution of Relative Humidity.
\newblock \emph{Journal of the Atmospheric Sciences}, 24(3):241–259, 1967.
\newblock \doi{10.1175/1520-0469(1967)024<0241:teotaw>2.0.co;2}.

\bibitem[{Meinshausen et~al.(2017)Meinshausen, Vogel, Nauels, Lorbacher, Meinshausen et~al.}]{Mei2017}
Malte Meinshausen, Elisabeth Vogel, Alexander Nauels, Katja Lorbacher, Nicolai Meinshausen, et~al.
\newblock Historical Greenhouse Gas Concentrations for Climate Modelling ({{CMIP6}}).
\newblock \emph{Geoscientific Model Development}, 10(5):2057--2116, 2017.
\newblock \doi{10.5194/gmd-10-2057-2017}.

\bibitem[{Milinski et~al.(2020)Milinski, Maher, and Olonscheck}]{Milinski2020}
Sebastian Milinski, Nicola Maher, and Dirk Olonscheck.
\newblock How large does a large ensemble need to be?
\newblock \emph{Earth System Dynamics}, 11(4):885–901, 2020.
\newblock \doi{10.5194/esd-11-885-2020}.

\bibitem[{NOAA-GFDL(2024)}]{fregrid2024}
NOAA-GFDL.
\newblock {{NOAA-GFDL}}/{{FRE-NCtools}}.
\newblock NOAA - Geophysical Fluid Dynamics Laboratory, 2024.

\bibitem[{Perkins and Hakim(2021)}]{Perkins2021}
W.~A. Perkins and G.~J. Hakim.
\newblock Coupled Atmosphere–Ocean Reconstruction of the Last Millennium Using Online Data Assimilation.
\newblock \emph{Paleoceanography and Paleoclimatology}, 36(5), 2021.
\newblock \doi{10.1029/2020pa003959}.

\bibitem[{Price et~al.(2023)Price, Sanchez-Gonzalez, Alet, Andersson, El-Kadi et~al.}]{Price2023}
Ilan Price, Alvaro Sanchez-Gonzalez, Ferran Alet, Tom~R. Andersson, Andrew El-Kadi, et~al.
\newblock GenCast: Diffusion-based ensemble forecasting for medium-range weather.
\newblock 2023.
\newblock \doi{10.48550/ARXIV.2312.15796}.

\bibitem[{Price(1981)}]{Price1981}
James~F. Price.
\newblock Upper Ocean Response to a Hurricane.
\newblock \emph{Journal of Physical Oceanography}, 11(2):153–175, 1981.
\newblock \doi{10.1175/1520-0485(1981)011<0153:uortah>2.0.co;2}.

\bibitem[{Rasp et~al.(2024)Rasp, Hoyer, Merose, Langmore, Battaglia et~al.}]{Rasp2024}
Stephan Rasp, Stephan Hoyer, Alexander Merose, Ian Langmore, Peter Battaglia, et~al.
\newblock WeatherBench 2: A Benchmark for the Next Generation of Data‐Driven Global Weather Models.
\newblock \emph{Journal of Advances in Modeling Earth Systems}, 16(6), 2024.
\newblock \doi{10.1029/2023ms004019}.

\bibitem[{R\"{u}hling~Cachay et~al.(2024)R\"{u}hling~Cachay, Henn, Watt-Meyer, Bretherton, and Yu}]{Salva2024}
Salva R\"{u}hling~Cachay, Brian Henn, Oliver Watt-Meyer, Christopher~S. Bretherton, and Rose Yu.
\newblock Probabilistic Emulation of a Global Climate Model with Spherical DYffusion.
\newblock 2024.
\newblock \doi{10.48550/ARXIV.2406.14798}.

\bibitem[{Russell and Kertész(2017)}]{metview}
Iain Russell and Sandor Kertész.
\newblock Metview.
\newblock 2017.

\bibitem[{Screen et~al.(2012)Screen, Deser, and Simmonds}]{Screen2012}
J.~A. Screen, C.~Deser, and I.~Simmonds.
\newblock Local and remote controls on observed Arctic warming.
\newblock \emph{Geophysical Research Letters}, 39(10), 2012.
\newblock \doi{https://doi.org/10.1029/2012GL051598}.

\bibitem[{Solomon(1999)}]{Solomon1999}
Susan Solomon.
\newblock Stratospheric ozone depletion: A review of concepts and history.
\newblock \emph{Reviews of Geophysics}, 37(3):275–316, 1999.
\newblock \doi{10.1029/1999rg900008}.

\bibitem[{Taylor et~al.(2000)Taylor, Williamson, and Francis}]{Tay2000}
Karl~E. Taylor, David Williamson, and Zwiers Francis.
\newblock The {{Sea Surface Temperature}} and {{Sea-Ice Concentration Boundary Conditions}} for {{AMIP II Simulations}}.
\newblock Technical report, Lawrence Livermore National Laboratory, 2000.

\bibitem[{Trenberth(1997)}]{Trenberth1997}
Kevin~E. Trenberth.
\newblock The Definition of El Niño.
\newblock \emph{Bulletin of the American Meteorological Society}, 78(12):2771–2777, 1997.
\newblock \doi{10.1175/1520-0477(1997)078<2771:tdoeno>2.0.co;2}.

\bibitem[{Trenberth et~al.(2011)Trenberth, Fasullo, and Mackaro}]{Trenberth2011}
Kevin~E. Trenberth, John~T. Fasullo, and Jessica Mackaro.
\newblock Atmospheric Moisture Transports from Ocean to Land and Global Energy Flows in Reanalyses.
\newblock \emph{Journal of Climate}, 24(18):4907–4924, 2011.
\newblock \doi{10.1175/2011jcli4171.1}.

\bibitem[{Ullrich et~al.(2021)Ullrich, Zarzycki, McClenny, Pinheiro, Stansfield et~al.}]{Ullrich2021}
Paul~A. Ullrich, Colin~M. Zarzycki, Elizabeth~E. McClenny, Marielle~C. Pinheiro, Alyssa~M. Stansfield, et~al.
\newblock TempestExtremes v2.1: a community framework for feature detection, tracking, and analysis in large datasets.
\newblock \emph{Geoscientific Model Development}, 14(8):5023–5048, 2021.
\newblock \doi{10.5194/gmd-14-5023-2021}.

\bibitem[{Urraca et~al.(2018)Urraca, Huld, Gracia-Amillo, Martinez-de Pison, Kaspar et~al.}]{Urraca2018}
Ruben Urraca, Thomas Huld, Ana Gracia-Amillo, Francisco~Javier Martinez-de Pison, Frank Kaspar, et~al.
\newblock Evaluation of global horizontal irradiance estimates from ERA5 and COSMO-REA6 reanalyses using ground and satellite-based data.
\newblock \emph{Solar Energy}, 164:339–354, 2018.
\newblock \doi{10.1016/j.solener.2018.02.059}.

\bibitem[{Vecchi et~al.(2021)Vecchi, Landsea, Zhang, Villarini, and Knutson}]{Vecchi2021}
Gabriel~A. Vecchi, Christopher Landsea, Wei Zhang, Gabriele Villarini, and Thomas Knutson.
\newblock Changes in Atlantic major hurricane frequency since the late-19th century.
\newblock \emph{Nature Communications}, 12(1), 2021.
\newblock \doi{10.1038/s41467-021-24268-5}.

\bibitem[{Waliser et~al.(2009)}]{Waliser2009}
D.~Waliser et~al.
\newblock {MJO} Simulation Diagnostics.
\newblock \emph{Journal of Climate}, 22(11):3006–3030, 2009.
\newblock \doi{10.1175/2008jcli2731.1}.

\bibitem[{Waliser et~al.(2003)Waliser, Jin, Kang, Stern, Schubert et~al.}]{Waliser2003}
D.~E. Waliser, K.~Jin, I.-S. Kang, W.~F. Stern, S.~D. Schubert, et~al.
\newblock AGCM simulations of intraseasonal variability associated with the Asian summer monsoon.
\newblock \emph{Climate Dynamics}, 21(5–6):423–446, 2003.
\newblock \doi{10.1007/s00382-003-0337-1}.

\bibitem[{Watson‐Parris et~al.(2022)Watson‐Parris, Rao, Olivié, Seland, Nowack et~al.}]{WatsonParris2022}
D.~Watson‐Parris, Y.~Rao, D.~Olivié, Ø. Seland, P.~Nowack, et~al.
\newblock ClimateBench v1.0: A Benchmark for Data‐Driven Climate Projections.
\newblock \emph{Journal of Advances in Modeling Earth Systems}, 14(10), 2022.
\newblock \doi{10.1029/2021ms002954}.

\bibitem[{Watt-Meyer et~al.(2023)Watt-Meyer, Dresdner, McGibbon, Clark, Henn et~al.}]{WattMeyer2023}
Oliver Watt-Meyer, Gideon Dresdner, Jeremy McGibbon, Spencer~K. Clark, Brian Henn, et~al.
\newblock {ACE: A} fast, skillful learned global atmospheric model for climate prediction.
\newblock 2023.
\newblock \doi{10.48550/arxiv.2310.02074}.

\bibitem[{Watt‐Meyer et~al.(2024)Watt‐Meyer, Brenowitz, Clark, Henn, Kwa et~al.}]{WattMeyer2024}
Oliver Watt‐Meyer, Noah~D. Brenowitz, Spencer~K. Clark, Brian Henn, Anna Kwa, et~al.
\newblock Neural Network Parameterization of Subgrid‐Scale Physics From a Realistic Geography Global Storm‐Resolving Simulation.
\newblock \emph{Journal of Advances in Modeling Earth Systems}, 16(2), 2024.
\newblock \doi{10.1029/2023ms003668}.

\bibitem[{Weyn et~al.(2020)Weyn, Durran, and Caruana}]{Weyn2020}
Jonathan~A. Weyn, Dale~R. Durran, and Rich Caruana.
\newblock Improving Data‐Driven Global Weather Prediction Using Deep Convolutional Neural Networks on a Cubed Sphere.
\newblock \emph{Journal of Advances in Modeling Earth Systems}, 12(9), 2020.
\newblock \doi{10.1029/2020ms002109}.

\bibitem[{Wheeler and Kiladis(1999)}]{Wheeler1999}
Matthew Wheeler and George~N. Kiladis.
\newblock Convectively Coupled Equatorial Waves: Analysis of Clouds and Temperature in the Wavenumber–Frequency Domain.
\newblock \emph{Journal of the Atmospheric Sciences}, 56(3):374–399, 1999.
\newblock \doi{10.1175/1520-0469(1999)056<0374:ccewao>2.0.co;2}.

\bibitem[{Zhang(2005)}]{Zhang2005}
Chidong Zhang.
\newblock Madden‐Julian Oscillation.
\newblock \emph{Reviews of Geophysics}, 43(2), 2005.
\newblock \doi{10.1029/2004rg000158}.

\bibitem[{Zhou et~al.(2022)Zhou, Harris, Chen, Gao, Guo et~al.}]{Zho2022}
Linjiong Zhou, Lucas Harris, Jan-Huey Chen, Kun Gao, Huan Guo, et~al.
\newblock Improving {{Global Weather Prediction}} in {{GFDL SHiELD Through}} an {{Upgraded GFDL Cloud Microphysics Scheme}}.
\newblock \emph{Journal of Advances in Modeling Earth Systems}, 14(7):e2021MS002971, 2022.
\newblock \doi{10.1029/2021MS002971}.

\end{thebibliography}
\bibliographystyle{main}

\newpage

\appendix

\section{Supplemental results}

\subsection{Spatial variability of temperature trends}
\label{appendix:temp_trend_maps}
\begin{figure}[t]
    \centering
    \includegraphics[width=0.65\textwidth]{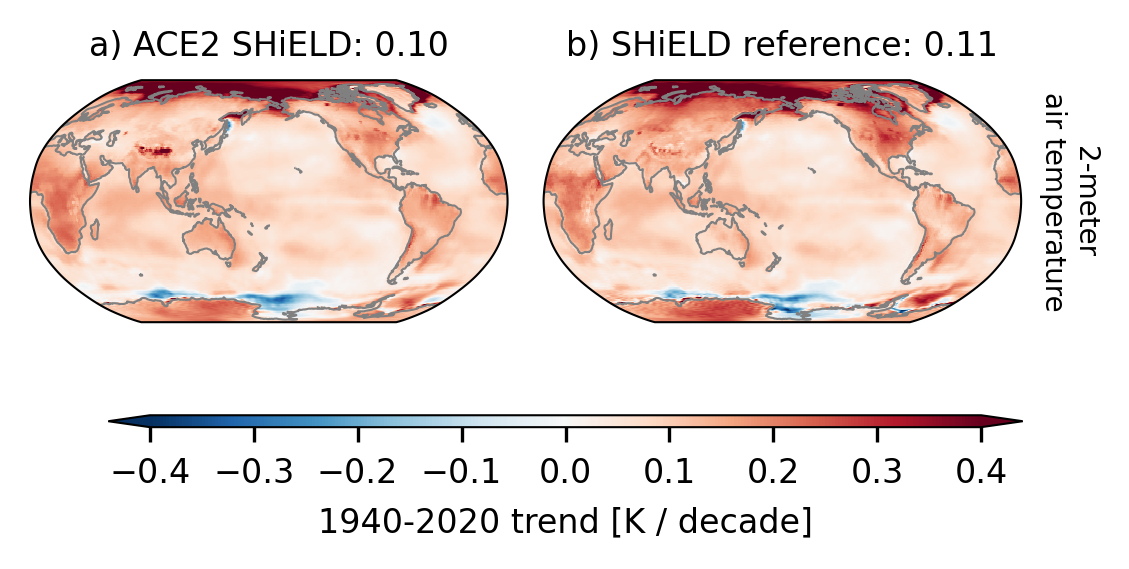}
    \caption{2-meter air temperature linear trend over 1940-2020 in (a) ACE2-SHiELD and (b) the reference SHiELD dataset. Trends shown are the average of trends computing in individual initial condition simulations: three for ACE2-SHiELD and two for SHiELD. Titles show the global-mean trend for each case in K / decade.}
    \label{fig:maps_2m_temperature_trends}
\end{figure}

Here we show maps of the trends in 2-meter air temperature over 1940-2020, the same period shown in Figure~\ref{fig:annual_mean_series}. Because sea surface temperature and sea-ice fraction are prescribed in ACE2 simulations, the emulator should predict the 2-meter air temperature to closely follow that of the forcing dataset over open ocean. Indeed, this is the case for ACE2-SHiELD (Fig.~\ref{fig:maps_2m_temperature_trends}). Furthermore, the pattern of surface air temperature warming over land and polar regions also closely follows the SHiELD reference over most regions. Some exceptions are the Himalaya, where ACE2-SHiELD shows too much warming, and Siberia and North America where ACE2-SHiELD shows slightly too little warming.

\subsection{OLR response to ENSO}
\label{appendix:OLR_ENSO}

\begin{figure}[t]
    \centering
    \includegraphics[width=\textwidth]{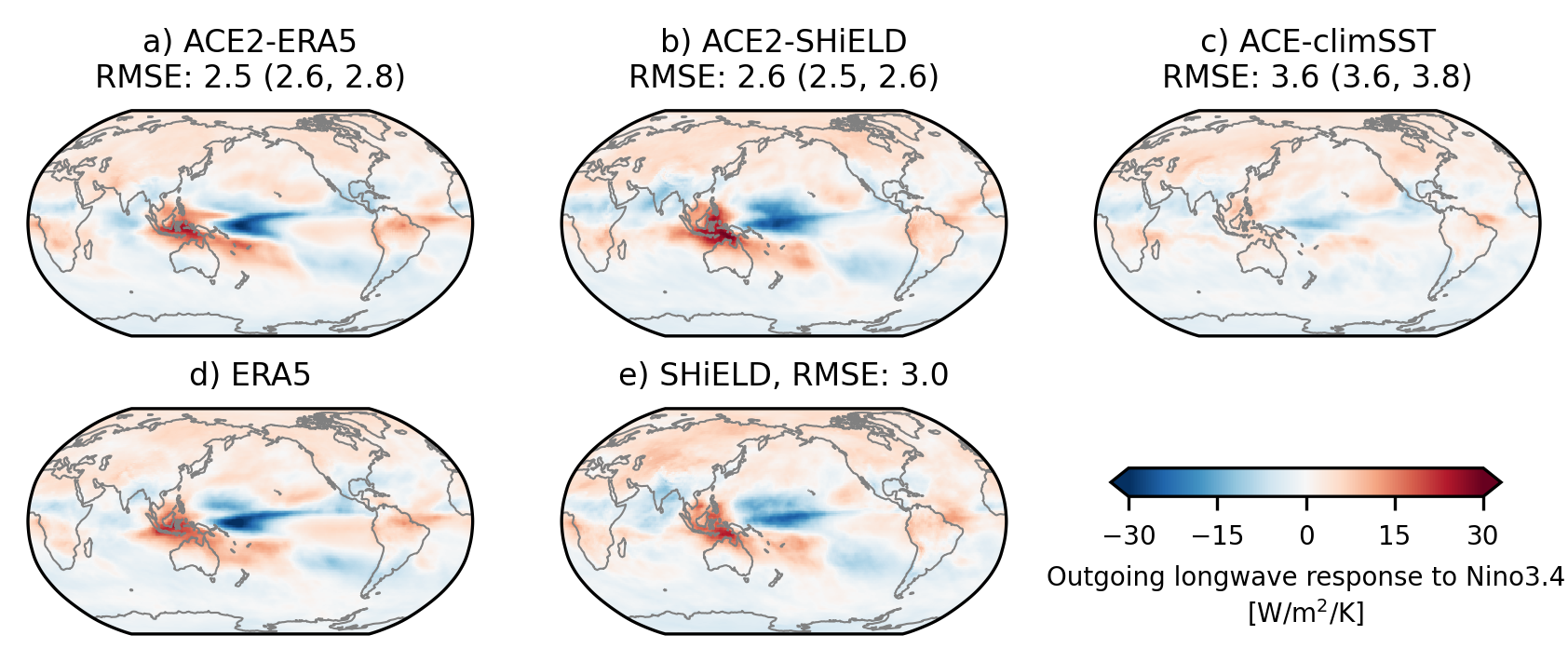}
    \caption{As in Fig \ref{fig:enso_precip}, but for outgoing longwave radiation at the top of atmosphere.}
    \label{fig:enso_olr}
\end{figure}

The predicted response of outgoing longwave radiation (OLR) to ENSO is shown in Fig. \ref{fig:enso_olr}. The mean error in ACE2's OLR response to Ni\~no3.4 is 2.6 W/m$^2$/K for both ACE2-ERA5 and -SHiELD, slightly lower than the reference variability of the response in SHiELD (3.0 W/m$^2$/K). ACE-climSST again has a muted OLR response over the tropical Pacific and a larger mean pattern error (3.7 W/m$^2$/K). 

\subsection{Tropical cyclone statistics and dependence on sea surface temperature dataset}
\begin{figure}[t]
    \centering
    \includegraphics[width=\textwidth]{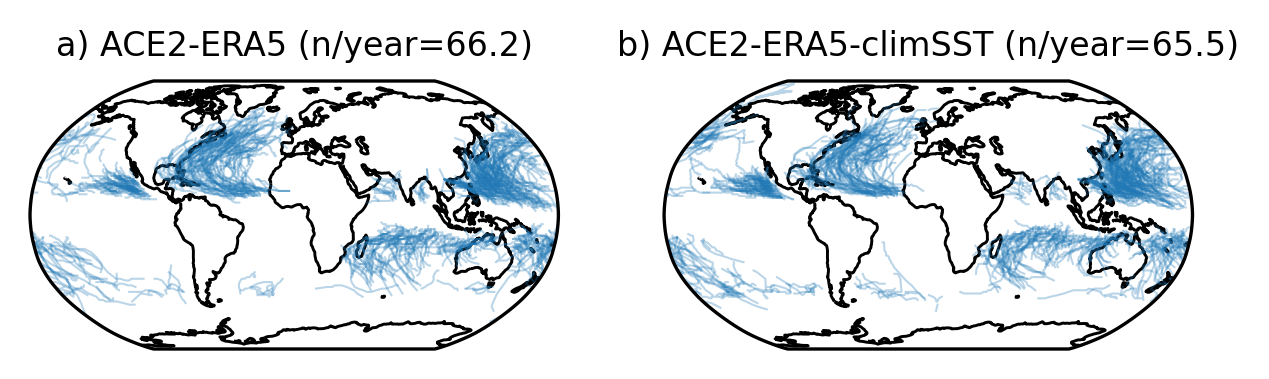}
    \caption{As in Figure~\ref{fig:tc-tracks} but comparing ACE2-ERA5 between 10-year runs forced by (left) observed historical sea surface temperature and sea ice fraction over the 2001-2010 period and (right) annually repeating 1990-2020 climatological sea surface temperature and sea ice fraction.}
    \label{fig:tc-climsst-tracks}
\end{figure}

As described in Section~\ref{subsec:tropical-cyclones}, a possible concern with our evaluation framework for evaluating tropical cyclones is that ACE2-ERA5 is forced with observed sea surface temperature, which will contain a signature of past tropical cyclones which can leave behind a cold wake \citep{Price1981}. Therefore, we run a simulation which forced by climatological sea surface temperature instead of using actual 2001-2010 sea surface temperatures. Reassuringly, we find that the total number of tropical cyclones per year and their geographic distribution is very similar between the two cases (Figure~\ref{fig:tc-climsst-tracks}), indicating that the use of true historical SSTs is not strongly influencing the generation of tropical cyclones in our test simulations.

\begin{figure}[t]
    \centering
    \includegraphics[width=0.85\textwidth]{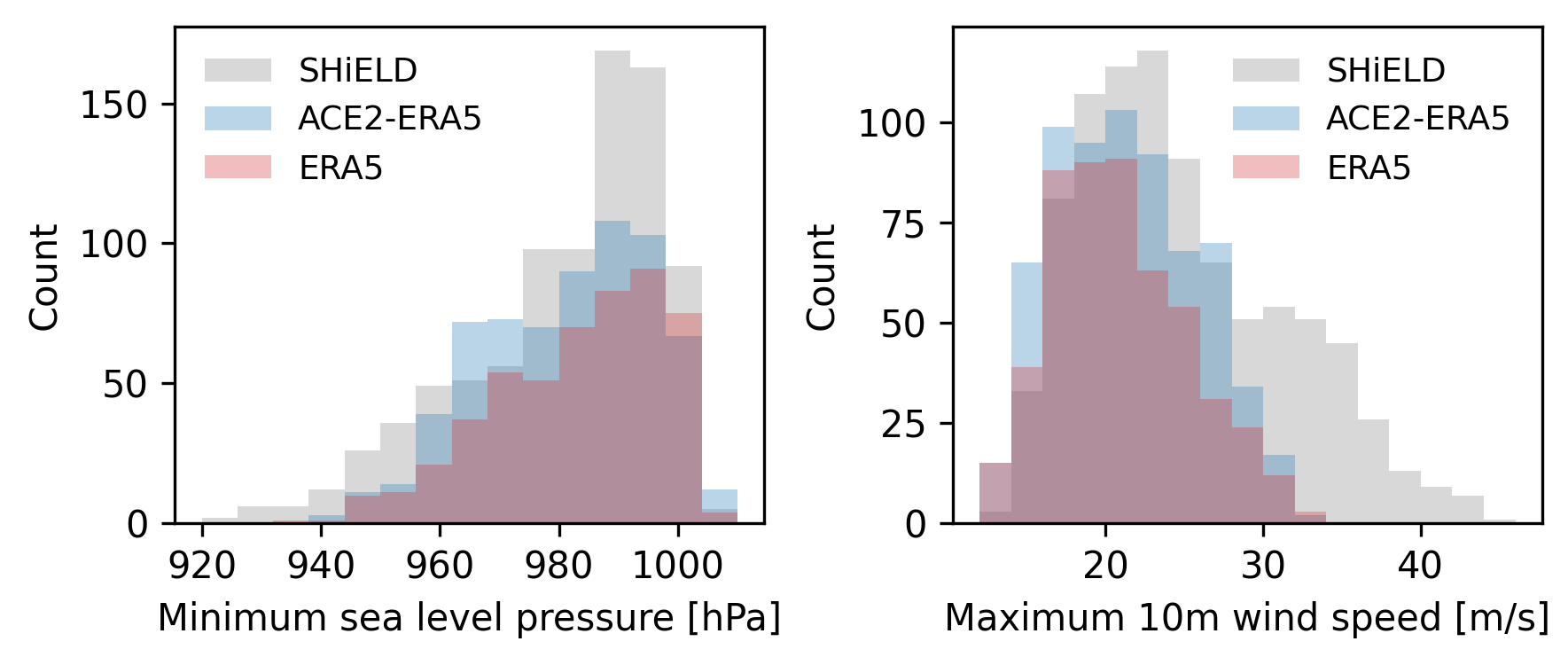}
    \caption{The (left) minimum sea-level pressure and (right) maximum 10m wind speed within 2$^{\circ}$ of the sea-level pressure minimum across all tropical cyclone tracks shown in Figure~\ref{fig:tc-tracks}.}
    \label{fig:tc-statistics}
\end{figure}

\subsection{Seed variability and checkpoint selection}
\label{appendix:discussion_seed_variability}
\begin{figure}[t]
    \centering
    \includegraphics[width=\textwidth]{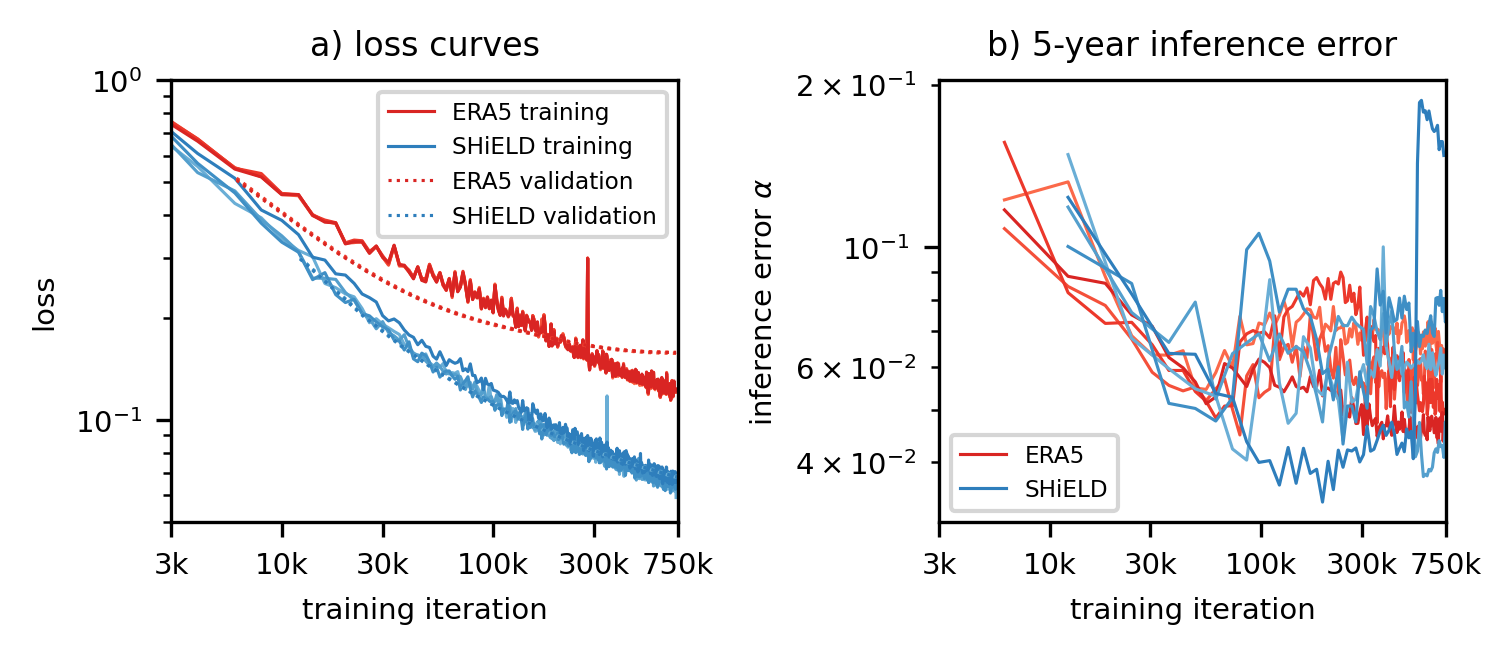}
    \caption{a) Training and validation loss for ACE2-ERA5 and ACE2-SHiELD, over a 4-member training ensemble. b) Inference error $\alpha$ (Equation~\ref{eq:time_mean_RMSE}) for the same training ensemble members, computed at the end of each epoch and averaged over eight 5-year long simulations initialized at evenly spaced intervals spanning 1996. Red lines are ACE2-ERA5 models, and blue lines are ACE2-SHiELD models. The ERA5 training had about 6,000 iterations per epoch, while the SHiELD training had about 12,000.}
    \label{fig:training_val_inference}
\end{figure}

As described in the ``Checkpoint selection based on climate skill'' paragraph in Section~\ref{subsec:training}, the loss function we use for training optimizes predictions over a 12-hour period (two 6-hour steps). Therefore, it is not necessarily expected that a model used autoregressively for many more steps (e.g. a 100-year simulation is about 146,000 steps) will necessarily be stable or accurate. In practice, using the SFNO architecture, hyperparameters and training setup described in this paper, we find that all models we train are indefinitely stable. However, their climate accuracy can vary significantly. Figure~\ref{fig:training_val_inference} demonstrates this by comparing the four models we trained each for the ERA5 and SHiELD datasets. Reassuringly, the training and validation losses are very similar across the training ensemble for each dataset and steadily decrease with more training (Fig.~\ref{fig:training_val_inference}a). Interestingly, the SHiELD and ERA5 datasets result in a final validation loss that is about two times larger for ACE2-ERA5 compared to ACE2-SHiELD. This suggests that the ERA5 dataset---whose production involves a data assimilation scheme and is designed to reproduce the true historical evolution of the atmosphere---is harder to learn than the SHiELD dataset, which involves learning the behavior of a 100$\,$km atmosphere-only numerical model. Regardless, Fig.~\ref{fig:training_val_inference}b shows that despite the steadily decreasing validation loss throughout training, the climate accuracy, measured as time-mean pattern RMSE averaged over output variables (see Equation~\ref{eq:time_mean_RMSE}), does not steadily improve with more training. Although it does so for the first few epochs (about 50,000 training iterations), after this point in training the climate accuracy of the model can deteriorate with more training, and then in some cases improve later on. Notably, the behavior of climate accuracy through training appears to be different for the SHiELD and ERA5 datasets. We note that some manual hyperparameter tuning was performed on the ERA5 dataset, and then the same hyperparameters were used for ACE2-SHiELD.

\begin{figure}[t]
    \centering
    \includegraphics[width=\textwidth]{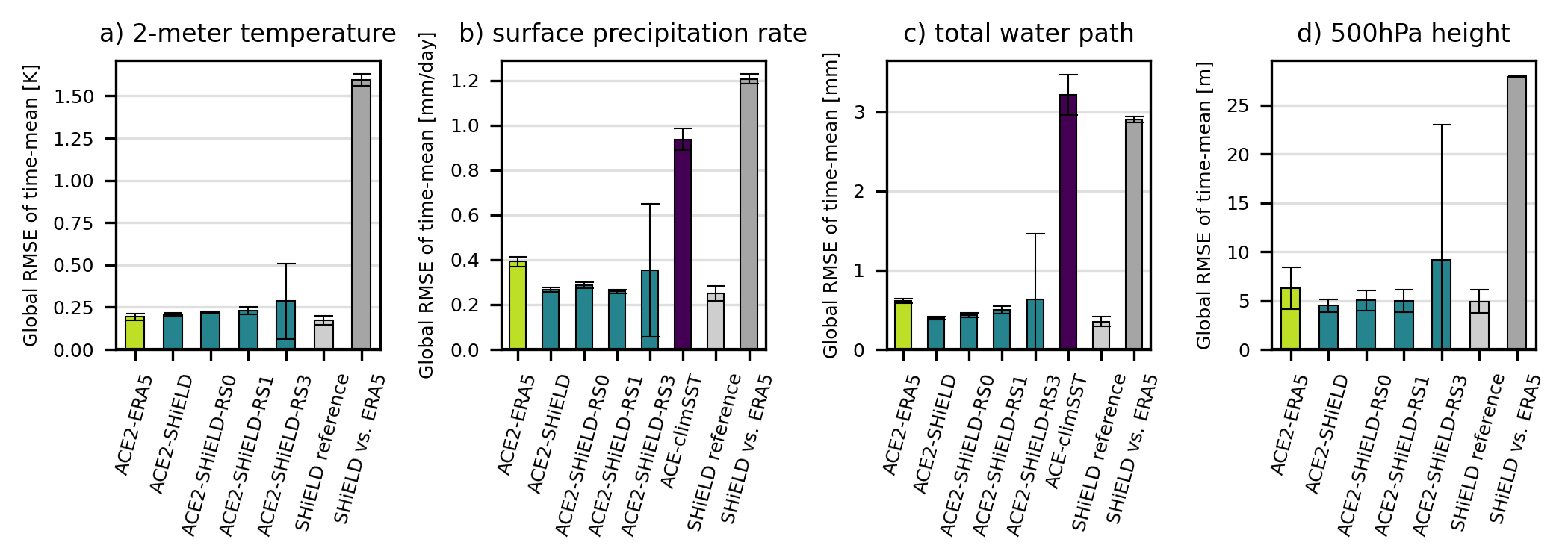}
    \caption{As in Figure~\ref{fig:time_mean_RMSE_10yr} but including all of the four models trained on the ACE2-SHiELD dataset and not including the NeuralGCM or ACE-climSST models.}
    \label{fig:time_mean_RMSE_with_ensemble}
\end{figure}

\subsection{Ablation of hard physical constraints}
In this section, we first demonstrate that the physical constraints described in Section~\ref{sec:constraints-methods} have the intended effect of resulting in closed global budgets of dry air mass and atmospheric moisture. We then examine their impact on the accuracy of long simulations. To show these effects, we train models with ``No constraints'' (specifically, removing the dry-air mass and moisture conservations, i.e. items 2-4 in Section~\ref{sec:constraints-methods}) and ``Dry air only'' (removing the moisture conservation, i.e. items 3 and 4 only). We compare these with the ACE2 model used in prior sections, which imposes both of these (``Dry air + moisture''). In all cases, the best model across random seed and epoch is chosen according to the checkpoint selection criteria described in Section~\ref{subsec:training}. However, to limit computational cost, we only run two random seeds for the ablated cases compared to four for the primary ``Dry air + moisture'' models. We do not show sensitivity to the positivity constraint (item 1 in Section~\ref{sec:constraints-methods} Hard Physical Constraints) but in practice found this had its desired effect without worsening any skill metric. For brevity, this section shows the ablations only for the model trained on the SHiELD dataset.

\begin{figure}[t]
    \centering
    \includegraphics[width=0.8\textwidth]{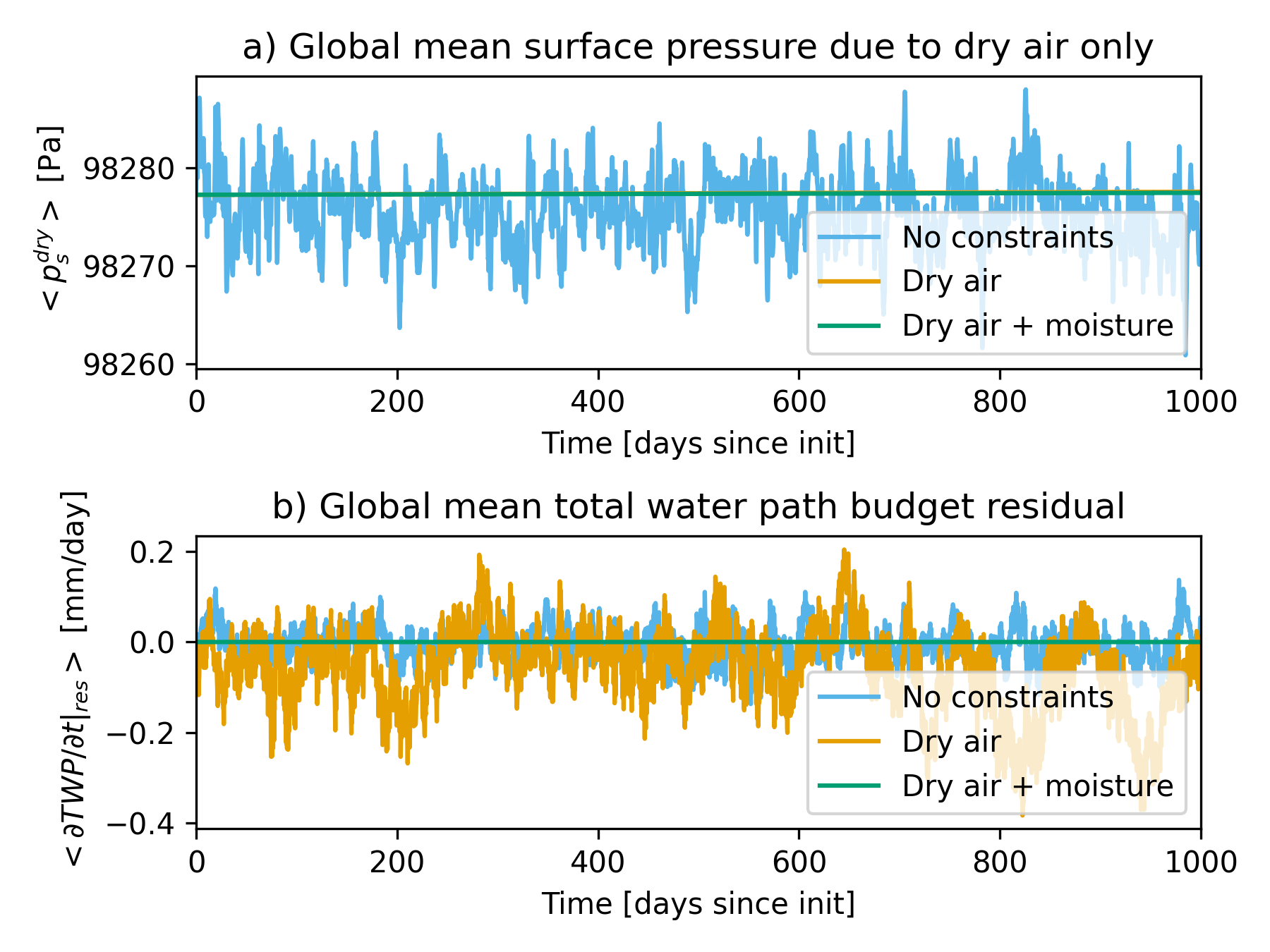}
    \caption{Timeseries of 6-hourly global mean (a) surface pressure due to dry air only and (b) total water path budget residual. The total water path budget residual is defined as the left-hand side minus the right-hand side of Equation~\ref{eq:columnmoisture}. In all cases, showing 1000 days of a single inference simulation initialized on 2001-01-01, although results are qualitatively simmilar throughout the run. "No constraints" is a model trained without dry air mass or moisture conservation imposed. ``Dry air" imposes only dry air mass conservation, while ``Dry air + moisture" imposes both. In (a) the orange and green lines are collocated. In all cases, ACE2-SHiELD models are shown.}
    \label{fig:physics-constraints-timeseries}
\end{figure}

Figure~\ref{fig:physics-constraints-timeseries} demonstrates that the imposed constraints have the desired effect. Conserving global dry air mass leads to a constant $\left <p_s^{dry} \right >$; without this constraint, the ACE2-SHiELD model predicts deviations of global dry air mass surface pressure of up to 20 Pa. In some prior cases, we have found that models without the dry air mass constraint have even larger biases in dry air mass (c.f. Figure 11 of \cite{WattMeyer2023}). Note that even with the dry air mass corrector applied, in some cases we find a very slight (<2Pa over 10 years) drift in global mean surface pressure due to dry air, possibly because of rounding errors when applying the very small correction at each 6-hour time step. The moisture constraint leads to a closed global total water path budget; without it, individual time step violations of this budget exceed 0.2 mm/day. 

\begin{figure}[t]
    \centering
    \includegraphics[width=\textwidth]{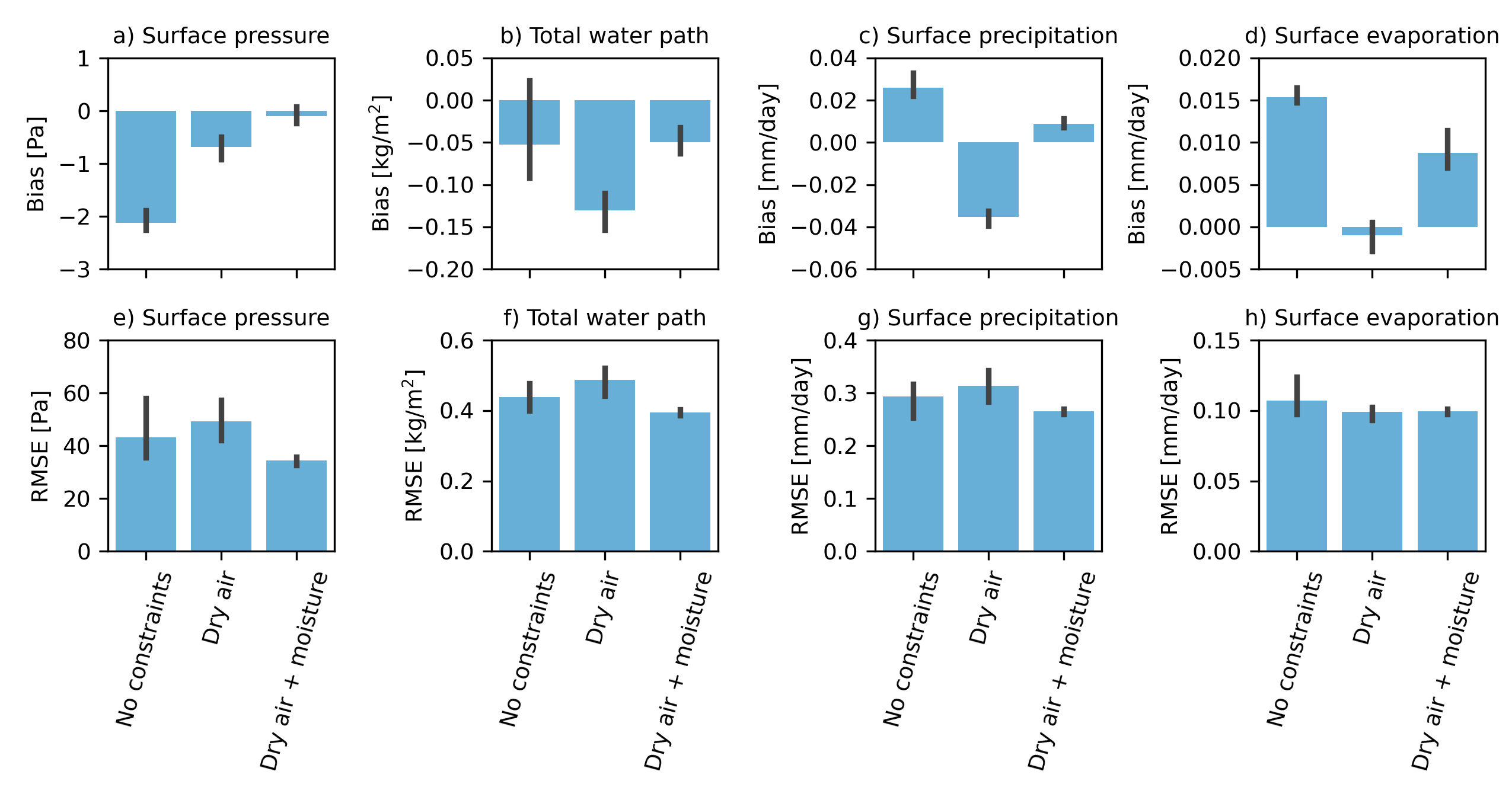}
    \caption{Across the three physical constraint ablation experiments with ACE2-SHiELD (top row) time-mean and global-mean bias of surface pressure, total water path, surface precipitation rate and surface evaporation rate and (bottom row) spatial RMSE of the time-mean pattern for the same four variables. The bars show the average bias over three 10-year simulations, initialized on 2001-01-01, 2001-01-02 and 2001-01-03. The error bars show the min/max range of over the three simulations.}
    \label{fig:physics-constraints-ablation-metrics}
\end{figure}

It is reassuring that the constraints have their intended effect, but it is also important to test how they affect the climate accuracy of ACE2. Imposing the dry air mass constraint robustly decreases the time- and global-mean bias surface pressure bias (Fig.~\ref{fig:physics-constraints-ablation-metrics}a). The total water path bias does not robustly change when adding the dry air mass or moisture constraints, and is small in all cases (Fig.~\ref{fig:physics-constraints-ablation-metrics}b). For the simulations shown here the precipitation and evaporation biases are reduced when imposing the moisture constraint (Fig.~\ref{fig:physics-constraints-ablation-metrics}c-d), but this was not true for all models trained. Given that the moisture constraint is based on the net flux of moisture through the surface, i.e. the difference between evaporation and precipitation, we do not \emph{a priori} expect the individual terms to be less biased with the constraint imposed.

For completeness, Figure~\ref{fig:physics-constraints-ablation-metrics}e-h  shows the time-mean pattern error for the same four variables. As expected given that we are imposing global mean constraints, the pattern errors are not significantly changed across the runs.

\subsubsection{Imposing constraints in ERA5, a non-conservative dataset}
\begin{figure}[t]
    \centering
    \includegraphics[width=0.9\textwidth]{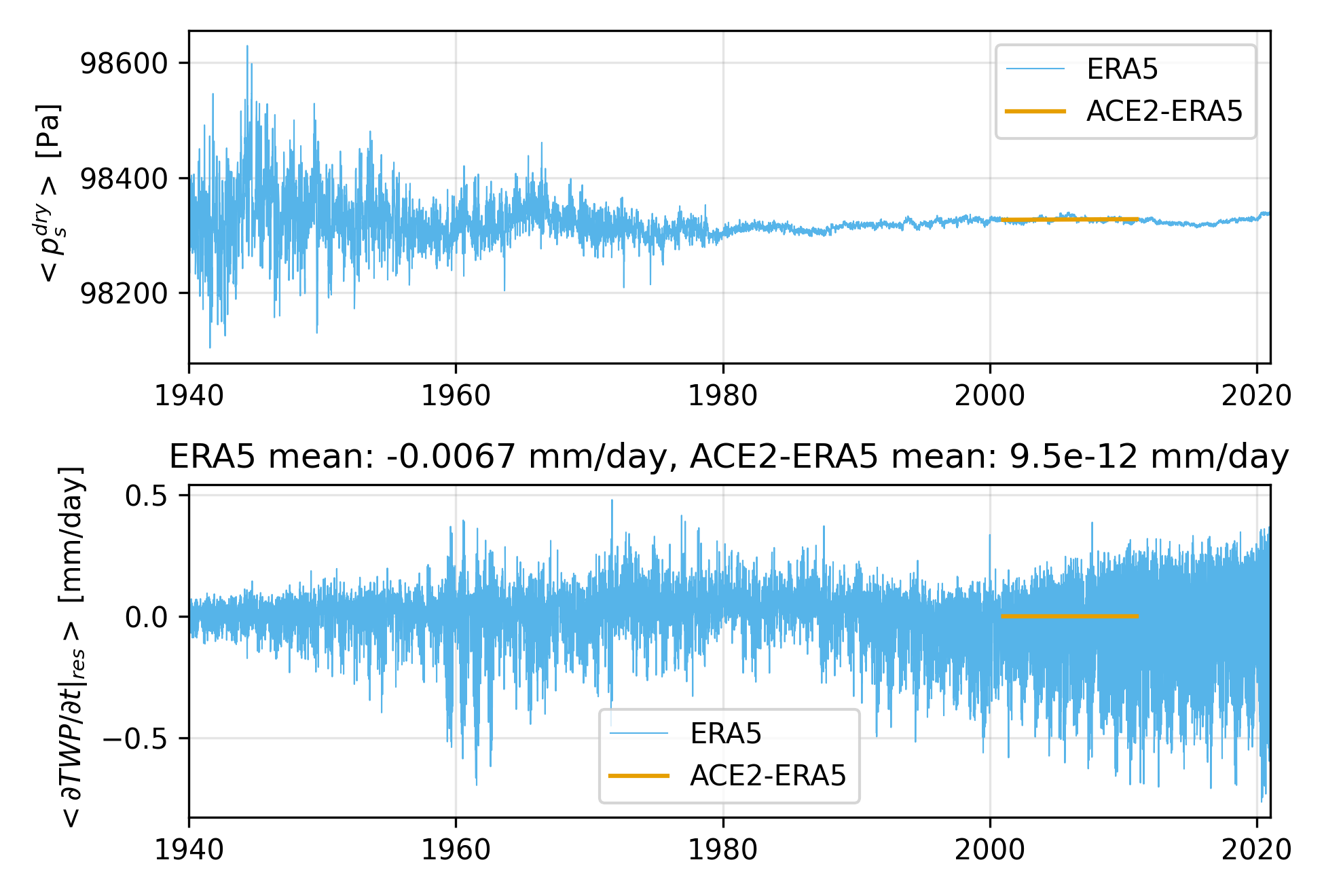}
    \caption{As in Figure~\ref{fig:physics-constraints-timeseries} but comparing the ERA5 dataset (blue) with a 10-year ACE2-ERA5 simulation initialized on 2001-01-01 (orange).}
    \label{fig:era5-budget}
\end{figure}

Reanalysis datasets generally do not have closed budgets because the increments from the data assimilation do not correspond to a particular physical process and will update the model state (e.g. specific humidity) without correspondingly updating the flux of that quantity into or out of the atmosphere (i.e. evaporation and precipitation) \citep{Trenberth2011}. Indeed, for the ERA5 reanalysis dataset, the global mean surface pressure due to dry air varies by up to 400 Pa in the earlier part of the data record when observing systems were more limited (Figure~\ref{fig:era5-budget} top row). After 1979, the fluctuations are much smaller, on the order of 10 Pa (c.f. Fig. 22 of \cite{Hersbach2020}). The ERA5 global moisture budget has residuals of up about 0.5mm/day on some time steps (Figure~\ref{fig:era5-budget} bottom row) but the time-mean of the residual is small (<0.005 mm/day). Despite these violations of expected physical budgets in the ERA5 dataset, we can still impose the the corrections described in Section~\ref{sec:constraints-methods}. Indeed, the ACE2-ERA5 emulator exactly obeys these budgets (orange lines in Figure~\ref{fig:era5-budget}). In practice, we find the time-mean biases of precipitation and evaporation are slightly larger in magnitude for ACE2-ERA5 compared to ACE2-SHiELD, but they are still only -0.05 mm/day and -0.04 mm/day respectively, small compared to the approximately 3 mm/day time- and global-mean of each of these variables.

\section{Dataset details}
\label{appendix:data}

The complete list of input and output variables used for ACE2 is given in Table~\ref{table:variables}. Variables which depend on height are labeled with a subscript $k$ which ranges form 0 to 7. The pressure at the interface between level $k$ and $k+1$ can be determined from the surface pressure and hybrid sigma-pressure coordinates \citep[e.g.][]{Collins2004} as:
\begin{equation}
\label{eq:akbk}
    p_k = a_k + b_k p_s
\end{equation}
where $a_k$ and $b_k$ (see Table~\ref{table:verticalcoord}) are coordinates chosen to approximately match the SPEEDY \citep{Kucharski2013} vertical coordinate.

\begin{table}
  \caption{Input and output variables for ACE2. The $k$ subscript refers to a vertical layer index, and ranges from 0 to 7 starting at the top of atmosphere and increasing towards the surface. The Time column indicates whether a variable represents the value at a particular time step (``Snapshot''), the average across the 6-hour time step (``Mean'') or a quantity which does not depend on time (``Invariant''). ``TOA'' denotes ``Top Of Atmosphere'', the climate model's upper boundary.}
  \label{table:variables}
  \centering
  \begin{tabular}{llll}
    \toprule
    \multicolumn{4}{c}{Prognostic (input and output)}   \\
    \midrule
    Symbol   & Description                                & Units & Time        \\
    \midrule
    $T_k$    & Air temperature                            & K     & Snapshot \\
    $q^T_k$   & Specific total water (vapor + condensates) & kg/kg & Snapshot \\
    $u_k$    & Windspeed in eastward direction            & m/s   & Snapshot \\
    $v_k$    & Windspeed in northward direction           & m/s   & Snapshot \\
    $T_s$     & Skin temperature of land or sea-ice            & K     & Snapshot \\
    $p_s$     & Atmospheric pressure at surface            & Pa    & Snapshot \\
    $T_{2m}$  & 2-meter air temperature                     & K    & Snapshot \\
    $q_{2m}$  & 2-meter specific humidity                     & kg/kg    & Snapshot \\
    $u_{10m}$  & 10-meter windspeed in eastward direction      & m/s    & Snapshot \\
    $v_{10m}$  &  10-meter windspeed in northward direction      & m/s    & Snapshot \\
    \midrule
    \multicolumn{4}{c}{Forcing (input only)}   \\
    \midrule
    Symbol       & Description                              & Units   & Time          \\
    \midrule
    DSWRF$_{toa}$ & Downward shortwave radiative flux at TOA & W/m$^2$ & Mean  \\
    $T_s$       & Skin temperature of open ocean           & K      & Snapshot \\
    $z_s$       & Surface height of topography             & m      & Invariant \\
    $f_l$       & Land grid cell fraction                  & $-$    & Invariant \\
    $f_o$       & Ocean grid cell fraction                 & $-$    & Snapshot \\
    $f_{si}$    & Sea-ice grid cell fraction               & $-$    & Snapshot \\
    $\mathrm{CO}_2$ & Global mean atmospheric carbon dioxide & ppm  & Snapshot \\
    \midrule
    \multicolumn{4}{c}{Diagnostic (output only)}   \\
    \midrule
    Symbol       & Description                                    & Units   & Time    \\
    \midrule
    USWRF$_{toa}$ & Upward shortwave radiative flux at TOA         & W/m$^2$ & Mean \\
    ULWRF$_{toa}$ & Upward longwave radiative flux at TOA          & W/m$^2$ & Mean  \\
    USWRF$_{sfc}$ & Upward shortwave radiative flux at surface     & W/m$^2$ & Mean  \\
    ULWRF$_{sfc}$ & Upward longwave radiative flux at surface      & W/m$^2$ & Mean  \\
    DSWRF$_{sfc}$ & Downward shortwave radiative flux at surface   & W/m$^2$ & Mean  \\
    DLWRF$_{sfc}$ & Downward longwave radiative flux at surface    & W/m$^2$ & Mean  \\
    $P$ & Surface precipitation rate (all phases)               & kg/m$^2$/s & Mean  \\
    $\left. \frac{\partial TWP}{\partial t}\right |_{adv}$ & Tendency of total water path from advection & kg/m$^2$/s & Mean \\
    $LHF$ & Surface latent heat flux                            & W/m$^2$ & Mean  \\
    $SHF$ & Surface sensible heat flux                          & W/m$^2$ & Mean  \\
    $Z_{500}$ & 500$\,$hPa geopotential height                    & m & Snapshot   \\
    $T_{850}$ & 850$\,$hPa air temperature                       & K & Snapshot   \\
    
    \bottomrule
  \end{tabular}
\end{table}




\begin{table}
  \caption{ACE2 vertical coordinate. Here $k$ indicates the vertical layer interface ranging from the top of the model's atmosphere $k=0$ to the Earth surface $k=8$. $a_k$ and $b_k$ define the vertical coordinate (see Equation~\ref{eq:akbk}). $I_k$ indicates what the corresponding vertical index is in the original dataset---79 layers for SHiELD and 137 layers for ERA5 (see \href{https://confluence.ecmwf.int/display/UDOC/L137+model+level+definitions}{https://confluence.ecmwf.int/display/UDOC/L137+model+level+definitions}). $p^{ref}_k$ is the pressure at model layer interfaces assuming $p_s=1000\,$hPa. To avoid interpolating across the native vertical layers when generating the 8-layer datasets, a slightly different vertical coordinate is used for the SHiELD and ERA5 datasets.}
  \label{table:verticalcoord}
  \centering
  \begin{tabular}{l|llll|llll}
    \toprule
     & \multicolumn{4}{c|}{SHiELD} & \multicolumn{4}{c}{ERA5}\\
    \midrule
    $k$     & $a_k$ [Pa]  & $b_k$ [-] & $I_k$ & $p^{ref}_k$ [hPa] & $a_k$ [Pa] & $b_k$ [-] & $I_k$ & $p^{ref}_k$ [hPa] \\
    \midrule
    0       & 300.0     & 0.0        & 0     & 3.0    & 0.0      & 0.0        &  0   &  0      \\
    1       & 5247.94   & 0.0        & 11    & 52.5  & 5119.90   & 0.0        &  48  &  51.2   \\
    2       & 12990.4   & 0.0090007  & 21    & 139   & 13881.3   & 0.00537781 &  67  &  144    \\
    3       & 14738.1   & 0.113764   & 30    & 261   & 19343.5   & 0.0597284  &  79  &  253    \\
    4       & 12854.1   & 0.286445   & 39    & 415   & 20087.1   & 0.203491   &  90  &  404    \\
    5       & 9156.07   & 0.510134   & 49    & 601   & 15596.7   & 0.438391   &  100 &  594    \\
    6       & 5484.37   & 0.710333   & 58    & 765   & 8880.45   & 0.680643   &  109 &  769    \\
    7       & 2261.67   & 0.879658   & 67    & 902   & 3057.27   & 0.873929   &  119 &  905    \\
    8       & 0.0       & 1.0        & 79    & 1000  & 0.0       & 1.0        &  137 &  1000   \\
    \bottomrule
  \end{tabular}
\end{table}


\section{Training hyperparameters}
\label{appendix:hyperparameters}
Table~\ref{table:sfnoparams} lists the SFNO hyperparameters used in this study. See \cite{Bonev2023} for details about the meaning of these parameters. The only modification to the architecture of SFNO made in this work is in the first spherical harmonic transform and the last inverse spherical harmonic transform, where Gauss-Legendre quadrature is used, as our data is on the Gaussian grid as opposed to the equiangular latitude-longitude grid used in \cite{Bonev2023} (see horizontal regridding section of Appendix~\ref{appendix:data}).

Table~\ref{table:optimizationparams} lists the hyperparameters used for optimization. Model parameters were averaged across training step using an exponential moving average (EMA).

\begin{table}
  \caption{SFNO hyperparameters. Names correspond to the definition of the SphericalFourierNeuralOperatorNet class found here: \url{https://github.com/ai2cm/ace/blob/06d145df7bca712f3957d2eaabc20e9b87a4d207/fme/fme/ace/models/modulus/sfnonet.py\#L255}. All configuration options not listed here are set to the defaults at the linked code.}
  \label{table:sfnoparams}
  \centering
  \begin{tabular}{ll}
    \toprule
    Name     & Value \\
    \midrule
    \texttt{embed\char`_dim}        & 384       \\
    \texttt{filter\char`_type}      & linear    \\
    \texttt{num\char`_layers}       & 8         \\
    \texttt{operator\char`_type}    & dhconv    \\
    \texttt{scale\char`_factor}     & 1         \\
    \texttt{spectral\char`_layers}  & 3         \\
    \bottomrule
  \end{tabular}
\end{table}

\begin{table}
  \caption{Optimization hyperparameters.}
  \label{table:optimizationparams}
  \centering
  \begin{tabular}{ll}
    \toprule
    Name     & Value \\
    \midrule
    Optimizer                        & AdamW       \\
    Weight decay                     & 0.01        \\
    Initial learning rate            & $1 \times 10^{-4}$    \\
    Learning rate schedule           & None        \\
    Number of epochs                 & 120 (ERA5 dataset), 60 (SHiELD dataset)  \\
    Batch size                       & 16          \\
    Exponential moving average decay rate             & 0.999 \\
    \bottomrule
  \end{tabular}
\end{table}

\end{document}